# Title: Exoplanet Biosignatures: Observational Prospects
Short title: Observational Prospects for Biosignatures

Yuka Fujii[1,2], Daniel Angerhausen[3,4], Russell Deitrick[5,6],
Shawn Domagal-Goldman[6,7], John Lee Grenfell[8], Yasunori Hori[9],
Stephen R. Kane[10], Enric Palle[11,12], Heike Rauer[8,13], Nicholas Siegler[14,15],
Karl Stapelfeldt[14,15], Kevin B. Stevenson[16]

[1]NASA Goddard Institute for Space Studies, 2880 Broadway, New York, NY
[2]Earth-Life Science Institute, Tokyo Institute of Technology, Ookayama, Meguro, Tokyo 152-8550, Japan
[3]CSH Fellow for Exoplanetary Astronomy, Center for Space and Habitability (CSH), Universität Bern, Sidlerstrasse 5, 3012 Bern, Switzerland
[4]Blue Marble Space Institute of Science, 1001 4th ave, Suite 3201Seattle, Washington 98154 USA
[5]Department of Astronomy, University of Washington, 3910 15th Ave NE, Seattle, WA, USA
[6]NASA Astrobiology Institute's Virtual Planetary Laboratory
[7]NASA Goddard Space Flight Center, Greenbelt, MD, USA
[8]Department of Extrasolar Planets and Atmospheres (EPA), Institute of Planetary Research, German Aerospace Centre (DLR), Rutherfordstraße 2, 12489 Berlin, Germany
[9]Astrobiology Center, National Institutes of Natural Sciences (NINS), 2-21-1 Osawa, Mitaka, Tokyo 1818588, Japan
[10]Department of Earth Sciences, University of California, Riverside, CA 92521, USA
[11]Instituto de Astrofísica de Canarias, Vía Láctea s/n, E-38205 La Laguna, Tenerife, Spain
[12]Departamento de Astrofísica, Universidad de La Laguna, Spain
[13]Center for Astronomy and Astrophysics, Berlin Institute of Technology, Hardenbergstrasse 36, 10623 Berlin, Germany
[14]Jet Propulsion Laboratory, California Institute of Technology, Pasadena, CA, USA
[15]NASA Exoplanet Exploration Office
[16]Space Telescope Science Institute, Baltimore, MD 21218, USA

Corresponding author: Yuka Fujii
Earth-Life Science Institute, Tokyo Institute of Technology
2-12-1 I7E-307 Ookayama, Meguro, Tokyo 152-8550, Japan
tel: +81-3-5734-2802  fax: + 81-3-5734-3416
email: yuka.fujii@elsi.jp

## Abstract

Exoplanet hunting efforts have revealed the prevalence of exotic worlds with diverse properties, including Earth-sized bodies, which has fueled our endeavor to search for life beyond the Solar System. Accumulating experiences in astrophysical, chemical, and climatological characterization of uninhabitable planets are paving the way to characterization of potentially habitable planets. In this paper, we review our possibilities and limitations in characterizing temperate terrestrial planets with future observational capabilities through 2030s and beyond, as a basis of a broad range of discussions on how to advance "astrobiology" with exoplanets. We discuss the observability of not only the proposed biosignature candidates themselves, but also of more general planetary properties that provide circumstantial evidence, since the evaluation of any biosignature candidate relies on their context. Characterization of temperate Earth-size planets in the coming years will focus on those around nearby late-type stars. JWST and later 30 meter-class ground-based telescopes will empower their chemical investigations. Spectroscopic studies of potentially habitable planets around solar-type stars will likely require a designated spacecraft mission for direct imaging, leveraging technologies that are already being developed and tested as part of the WFIRST mission. Successful initial characterization of a few nearby targets will be an important touchstone toward a more detailed scrutiny and a larger survey that are envisioned beyond 2030. The broad outlook this paper presents may help develop new observational techniques to detect relevant features as well as frameworks to diagnose planets based on the observables.

## Table of Contents

# 1. Introduction

In the endeavor to discover life beyond the Solar System, the most critical step is to detect photometric, spectroscopic, and/or polarimetric properties of "potentially habitable exoplanets" and search for features related to life. The ways in which such observations can be utilized to detect life at various confidence levels are described in the other manuscripts in this issue (Schwieterman et al., 2017; Meadows et al., 2017; Catling et al., 2017; Walker et al., 2017). The idea of building a space-based direct-imaging observatory specifically aimed at detecting signs of life on Earth-like planets dates back to the 1990s (e.g., Burke 1992; Elachi et al. 1996), which elicited the *Terrestrial Planet Finder* (*TPF*) mission studies by the National Aeronautics and Space Administration (NASA) (Beichman et al. 1999; Lawson et al. 2007; Levine et al. 2009) and *Darwin* mission concepts of the European Space Agency (ESA) (Léger et al. 1996; Fridlund et al. 2000). The *Advanced Technology Large Aperture Space Telescope* (*ATLAST*) concept represents a general-purpose observatory capable of exoplanet direct-imaging with even larger apertures (Postman et al. 2009), and was later updated as the *High Definition Space Telescope* (HDST, AURA, http://www.hdstvision.org/). While the last Astrophysics Decadal Survey of the United States did not prioritize any of these concepts (https://www.nap.edu/catalog/12951/new-worlds-new-horizons-in-astronomy-and-astrophysics), it did recommend exoplanet technology development as its top medium-class investment.

Since these early mission studies, a huge expansion of exoplanet science has taken place thanks to discoveries and initial characterization made by radial velocity, transit, microlensing surveys, transit spectroscopy of close-in planets, and direct imaging of uninhabitable self-luminous exoplanets. These observations have revealed thousands of exoplanets, allowing for analyses of demographic trends in the exoplanet population. Of particular interest from an astrobiological viewpoint is the occurrence rate of terrestrial planets in so-called Habitable Zones (HZs), i.e., the circumstellar region in which liquid water could exist on the surface of a terrestrial planet (Kasting et al. 1993). This rate is conventionally represented by $\eta_\oplus$ and estimates were obtained employing various criteria for the "terrestrial" size and for the range of HZs (Catanzarite & Shao 2011; Traub 2012; Bonfils et al. 2013; Gaidos 2013; Dressing & Charbonneau 2013, 2015; Kopparapu 2013; Petigura et al. 2013; Morton & Swift 2014; Silburt et al. 2015; Zsom 2015; Burke et al. 2015). While the estimates span a range from a few percent up to the order of unity reflecting the differences in the datasets and the thresholds for the targets, it is now established that Earth-sized planets in HZs are not rare. Meanwhile, the analyses of the mass-radius relationship of close-in planets (period shorter than ~100 days) have revealed that most of the planets with radii below $1.5 - 2R_\oplus$ ($R_\oplus$ is the Earth radius) are consistent with rocky/metallic composition, while bigger planets have large scatter in bulk density with a substantial fraction of volatile-rich planets (e.g., Weiss & Marcy 2014; Rogers 2015; Kaltenegger 2017); Interestingly, a recent analysis indicates a gap in population between the planets smaller than $\sim 1.5R_\oplus$ and those larger than $\sim 2R_\oplus$ (Fulton et al. 2017). A few probably terrestrial planets around HZs have already been discovered in the Solar neighborhood: Proxima Centauri b, an Earth-mass planet receiving 65% of the incident flux received by the Earth, only 1.3 pc away (Anglada-Escudé et al. 2016); GJ 273 b, a planet a few times as massive as the Earth with an incident flux similar to that received by the Earth, 3.8 pc away (Astudillo-Defru et al. 2017b); seven transiting Earth-sized planets around an ultra-cool star TRAPPIST-1, 3-4 of which could conceivably be habitable, 12 pc away (Gillon et al. 2017); and LHS 1140 b, a large terrestrial planet, 12 pc away (Dittmann et al. 2017).

In parallel, substantial technological and methodological progress is being made through the characterization of larger and/or hotter exoplanets. Recently proven observational techniques to characterize planetary atmospheres include the usage of temporal variation to map the heterogeneity of planetary photospheric surfaces (e.g., Knutson et al. 2007; Knutson et al. 2012; Majeau et al. 2012; de Wit et al. 2012; Demory et al. 2013, 2016), and the usage of the cross-correlation analysis on high-resolution spectra to extract Doppler-shifted lines due to the planetary atmosphere (e.g., Snellen et al. 2010; Birkby et al. 2013; Konopacky et al. 2013). Lessons on data reduction processes and atmospheric retrieval techniques are being learned (see Deming & Seager 2017 for a review). Numerical simulations are also used to further develop the data analysis techniques of spectroscopic data (e.g., Line et al. 2013; Line & Parmentier 2016; Rocchetto et al. 2016; Deming & Sheppard 2017) and photometric light curves (see Cowan & Fujii 2017 for a review). The starlight suppression technologies for high-contrast imaging have been advanced by the successful ground-based direct-imaging observations using adaptive optics and coronagraphs (e.g., Kalas et al. 2008; Marois et al. 2008; Lagrange et al. 2010; Kuzuhara et al. 2013; Macintosh et al. 2015). Starshades have emerged as a viable alternative approach to coronagraphs (Cash 2006).

As these observations have progressed, theoretical work has been exploring the properties and diversity of temperate terrestrial planets, which could eventually be studied through similar techniques. The list of potential biosignatures continues to grow, and include the spectral features of atmospheric (volatile) molecules originated from possible life (e.g., $O_2$, $O_3$, $CH_4$, $N_2O$, $CH_3Cl$), the reflectance spectra of biological surfaces (e.g., vegetation's red edge, reflectance spectra of pigments), and the temporal variation of these signatures (see a review by Schwieterman et al., 2017 and references therein). It has also been recognized that the proposed potential biosignatures contain the risk of false positives, (i.e., they can be produced non-biologically under particular situations). Therefore, identifying an inhabited planet with confidence also requires as much contextual information as possible to evaluate the prospect of **a** non-biological origin of detected biosignature candidates and to find auxiliary evidence consistent with a biological origin (see Meadows et al., 2017 for a discussion on how $O_2$ would work as a biosignature, and see Catling et al., 2017 for a framework to assess potential biosignatures). The successful interpretation will require significant advances in our ability to model both inhabited and uninhabited worlds (see discussions in Walker et al., 2017) as well as the detailed observational data.

Founded on these ongoing observational, technological, and theoretical developments, new space-based missions and ground-based facilities will come into play in the near future. The planned new telescopes most relevant to the investigations of potentially habitable planets are listed in Table 1, together with their specifics and the expected usage. In this paper, we overview the capabilities of these future missions as well as the observational methods they will employ, and discuss what kind of properties of potentially habitable exoplanets could be observationally constrained. We do not intend to prioritize future projects or observational techniques. Instead, our aim is to share the ongoing efforts and limitations in exoplanet observations with a broad range of readers involved in astrobiology, so that we can be on the same page to think collaboratively about how to make the most of future opportunities to deduce useful information of planets.

In this paper, we use the term "potentially habitable exoplanets" to imply two properties of such planets: (1) terrestrial, i.e., inferred to have a well-defined surface and no voluminous gaseous envelope, and (2) in the HZ of their stars. For condition (1), we focus on small planets roughly up to $\sim 2R_\oplus$ in radius, and up to $\sim 10M_\oplus$ in mass where $M_\oplus$ is the Earth mass, referring to the recent observational evidence of the radius/mass range of planets consistent with no voluminous gaseous envelope (Kaltengger 2017). As for (2), to the first order the orbital distance of HZs scales with the square root of the stellar luminosity, and rough estimates tell that they are around 1AU for solar twins and around 0.01-0.3 AU for M-type stars ($M_\star \lesssim 0.5 M_\odot$, $L_\star \sim 10^{-4} \sim 10^{-1} L_\odot$ where $M_\star$ and $L_\star$ are stellar mass and luminosity, respectively, and $M_\odot$ and $L_\odot$ denote the solar values). Note that the exact location of the HZ of a given star depends on many factors including the stellar spectrum, planetary rotation, atmospheric properties, the initial amount of water and the evolutionary history of the surface environment (e.g., Kasting et al. 1993; Abe et al. 2011; Pierrehumbert 2011; Kopparapu et al. 2013, 2014, 2016; Leconte et al. 2013; Yang et al. 2013; Wolf and Toon 2014, 2015; Zsom et al. 2013; Kadoya and Tajika 2014; Ramirez and Kaltenegger 2014, 2016, 2017; Wolf et al. 2017). Outside HZs, planets may not necessarily be inhospitable; yet, their habitable environments are more likely to be confined to the subsurface (c.f. the "internal" oceans of Europa and Enceladus), and thus are probably more difficult to observe across interstellar distances.

The organization of this chapter is as follows. In Section 2, we broadly describe the overall trend in the exoplanet observations, which we expect to evolve from a focus on the astrophysical characterization of exoplanets toward their chemical, climatological, and astrobiological characterization. Then, we move on to how *individual* potentially habitable planets will be characterized from various aspects. We discuss transiting planets (Section 3) and planets with general orbital inclination (Section 4) separately, as the former enables some unique methods for astrophysical and chemical characterization and will be the prime targets in the coming few years. In each of these two sections, we review the methods for astronomical (mass, radius, orbit) and chemical/climatological (atmosphere, surface, etc.) characterizations, and the planned observational projects that may make use of the methods. While we try our best to reflect the state of the field at the time of writing, the specifics of the future missions are subject to change. Section 5 is devoted to how the contextual information, including host star properties and planetary system properties, will be obtained and how it will help in evaluating the planetary conditions. In Section 6, we introduce the mission concepts under development that envision commencing of operation beyond 2030, and explore more ambitious possibilities presented in the literature that could be planned further in the future. Lastly, Section 7 concludes this paper by placing the projects in a timeline, and discussing the work to be done.

## 2. From Astrophysical Characterization to Astrobiological Characterization

In this section, we briefly summarize how the exoplanet community as a whole plans to advance toward astrobiological investigations of exoplanets with future missions. This includes a description of discoveries our community anticipates occurring in three broad eras of exoplanet observations: (1) astrophysical characterization, (2) chemical/climatological characterization, and (3) astrobiological characterization. Astrobiological characterization can be seen as astrophysical, chemical, and climatological characterizations particularly for potentially habitable planets. Figure 1 is a schematic showing the relations among these regimes. The arrows of various missions reach in to the "astrobiological characterization" region to an extent reflecting approximately the similarity of their targets to the Earth; thus the missions primarily for potentially habitable planets around late-type stars are slightly shortened. This section can serve as a preview of the rest of the manuscript, in which we detail the methodologies and the specifics of the future projects to discuss how individual planets of astrobiological interest will be characterized.

## 2.1 The Era of Astrophysical Characterization of Exoplanets

We are in the golden age of the era focused on the detection and astrophysical characterization of exoplanets. After the pioneering *Convection, Rotation and planetary Transits* (*CoRoT*) mission demonstrated precision photometry from space, the larger aperture and higher photometric precision of the *Kepler* mission enabled thousands of planets to be discovered, including numerous Earth-sized planets. For many of these worlds, we have measured both size and mass, knowledge of which allows for inferences on the bulk composition of these planets. The large sample sizes have also allowed for trends to be uncovered in exoplanet populations. The combination of the exoplanet demographics and the simulations of bulk composition and density have led to the inference that, at least for close-in planets, there are three classes of planet size/mass: 1) planets with a rock-dominated composition that have small masses and radii; 2) planets with a gas-dominated composition that have large masses and radii; 3) and planets with intermediate sizes that have a composition which is dominated by neither rock nor gas.

The discoveries made during this era have also included multiple surprises in the orbital and size properties of planets. Hot Jupiters with large masses and short orbits were thought to be improbable if not impossible, yet were the first planets discovered around main sequence stars (Mayor & Queloz 1995). Circumbinary planets were proposed in science fiction lore yet were considered dubious by the astrophysics community until discovered by the *Kepler* mission (Doyle et al. 2011). The era of astrophysical characterization began with biases toward planets larger than Jupiter, and orbits shorter than Mercury's. Over time, detection techniques have improved to allow detections of planets with potentially habitable conditions. This began with discoveries by the *Kepler* mission (Borucki et al. 2011), has continued with ground-based surveys (e.g., Anglada-Escudé et al. 2012; Anglada-Escudé et al. 2016; Gillon et al. 2017; Astudillo-Defru et al. 2017b; Dittmann et al. 2017), and will continue further with ground-based measurements as well as the upcoming *TESS* and *PLATO* missions (Section 3.1.2). This "census" of astrophysical properties of exoplanets will be complemented well by the *Gaia* astrometric survey, which will be biased toward the detection of planets with orbits that extend beyond the Habitable Zone, and *WFIRST* microlensing surveys, which will be sensitive to the intermediate orbital regions. The latter will provide greater completeness to our survey of the abundance of potentially habitable worlds.
    .

## 2.2 The Era of Chemical Characterization of Exoplanets

Some of the recent exoplanet discoveries, and those anticipated from *TESS* and *CHEOPS*, will present the community with a pivot point to the next era of exoplanet characterization, which will be focused on chemical composition. Discoveries of nearby transiting worlds will enable follow-up transmission spectroscopy observations (Section 3.2). In principle the method is similar to the means by which transiting planets are discovered, but higher sensitivities allow transit events to be measured at multiple wavelengths. As the wavelength-dependence of the transit is a function of the atmosphere's opacity and scattering properties, this method will identify the chemical composition of exoplanet atmospheres. Further information can be obtained from spectroscopy of planetary eclipses, which extracts planetary dayside emissions (Section 3.3), and/or based on the phase curves of exoplanets, which probe the heterogeneity of the atmosphere and enable us to map some features (Section 4.2). While these kinds of chemical analyses have been done with the *Hubble Space Telescope* (*HST*) (e.g., Charbonneau et al. 2002; Vidal-Madjar et al. 2003; Kreidberg et al. 2014), *Spitzer Space Telescope* (e.g., Richardson et al. 2007; Grillmair et al. 2007; Knutson et al. 2007), *SOFIA* (e.g., Angerhausen et al. 2015) and ground-based observatories (e.g., Redfield et al. 2008), it will accelerate when the *James Webb Space Telescope* (*JWST*) launches (Section 3.2.3).

These observations by *JWST* should constitute the start of the golden era for the chemical characterization of exoplanets, which will continue with multiple observatories and techniques. *JWST* should have enough observation time for direct imaging of dozens of young gas giants, and should be able to measure their chemical inventories. The next generation of ground-based instruments (Section 3.1.2) will also enable detailed transmission spectroscopy of transiting gaseous planets and direct imaging observations of young gaseous planets in distant orbits. Ultimately, the 30 meter-class ground-based telescopes (also referred to as "Extremely Large Telescopes" or ELTs for short; Sections 3.2.3, 4.3.3, and 4.4.3) will eventually carry out more sensitive transit spectroscopy and direct imaging of many exoplanets, potentially down to sub-Neptune-sized planets. Future space missions dedicated to spectroscopy of exoplanets, *Fast Infrared Exoplanet Spectroscopy Survey Explorer* (FINESSE; Swain et al. 2012) and *Atmospheric Remote-sensing Infrared Exoplanet Large-survey* (ARIEL; Tinetti et al. 2016) plan to conduct a chemical survey of 500 and 1000 transiting planets, respectively, in the 2020s. The wealth of data on the atmospheric composition from these observatories will provide crucial insights into the formation histories of planetary systems, putting our own Solar System into a broader context.

## 2.3 The Era of Astrobiological Characterization of Exoplanets

Chemical characterization with *JWST* and the ELTs will also initiate the era of astrobiological characterization through a confirmation of habitable conditions and a search for signs of life on potentially habitable exoplanets. *JWST* should be capable of characterizing the atmospheric composition of at least one potentially Earth-like exoplanet (Stevenson et al. 2016), while the updated instruments with existing ground-based telescopes and future ELTs also plan to probe their atmospheres with transmission spectroscopy (Section 3.2), high-contrast imaging (Section 4.3), or high-contrast high-resolution observations (Section 4.4). The observations with these facilities will likely be limited to a few planets in orbit around cooler M-type stars. These stars are smaller than the Sun and have a relatively larger transit depth for an Earth-sized planet, and a planet-to-star contrast ratio that is a few orders of magnitude better than the contrast ratio of Earth-like planets to Sun-like stars. The habitability of such worlds has been brought into question based on complications stemming from the star's high-energy radiation (Ramirez & Kaltenegger 2014; Luger & Barnes 2015; Airapetian et al. 2017), and from the climate effects of synchronously rotating planets (Joshi et al. 1997; Joshi 2003; Wordsworth et al. 2011; Barnes et al. 2013). Regardless of the outcome, however, this will be the first time such observations are possible for temperate Earth-sized planets around other stars.

The golden age in the era of astrobiological characterization will likely require a space-based flagship mission that includes biosignature detection as a major design driver of the mission's architecture. Historically, the study of such missions has focused on direct imaging missions such as *TPF-C*, *TPF-I*, the *New Worlds Observer*, *THEIA*, and *Darwin*. However, biosignatures could also be detected via transit transmission/emission spectroscopy, if the observatory has sufficiently low noise characteristics. Currently, NASA is studying three flagship mission concepts in advance of the next U.S. Decadal survey, which all include a search for biosignatures in their design drivers: *Habitable Exoplanet Imaging Mission* (*HabEx*), *Large UltraViolet Optical and InfraRed surveyor* (*LUVOIR*), and *Origins Space Telescope* (*OST*) (Section 6.1). *OST* is a general mid-infrared observatory and is being designed with transit spectroscopy of potentially habitable worlds in mind; it will be more sensitive than *JWST*, and should extend observations to longer wavelengths and larger number of planets than what *JWST* can access. *LUVOIR* does not have transit spectroscopy as a central driver, but its large collecting area (> 8m mirror) should increase sensitivity, and do so at a wavelength range complementary to (shorter than) *JWST*'s. However, the primary targets of these transit spectroscopy observations would still be planets orbiting M-type stars. Both *HabEx* and *LUVOIR* aim to characterize terrestrial planets in the HZs around a variety of nearby stars, with most targets being F- G-, or K-type stars, via direct-imaging spectroscopy, and to conduct a range of general astrophysics observations that would place the exoplanet spectra in the context of the host star, comparative planetology, and cosmological history. They differ in their levels of quantitative ambition. These direct imaging observations would probe deeper into exoplanet atmospheres, at some wavelengths down to the surface. Thus, the missions designed with this technique in mind would be able to assess exoplanetary properties that will be difficult or impossible to otherwise observe. These three missions are discussed in more detail in Section 6.1.

The way individual planets of astrobiological interest are characterized will not necessarily proceed monotonically from astrophysical to chemical and climatological characterization. The possibilities to measure specific planetary properties depend on many factors, including whether the planets are transiting or not. Thus, we discuss prospects for transiting planets and planets with general orbital inclination separately in the following two sections. These prospects for future observations also depend on the spectral type of the host star. Therefore, we consider solar-type (F-, G-, K-type) stars and late-type (M-type) stars as two representative classes to give a rough idea. We summarize the prospects for each class (transiting or non-transiting planets around solar-type or late-type stars) in Table 2; Note that in reality the spectral type of the star is continuous and the scope of each observational method does not necessarily follow this classification strictly. Figure 2 summarizes some spectral signatures of interest, which will be discussed later in the paper.

## 3. Characterizing Transiting Planets

In this section, we focus on the methods to characterize potentially habitable planets that are applicable to transiting ones. Measurements of radius, mass, and the orbital elements are referred to as astrophysical characterization and discussed in Section 3.1. Chemical, climatological characterization through transmission spectroscopy and eclipse spectroscopy are discussed in Sections 3.2 and 3.3, respectively. In each of these subsections, we review the method(s), sensitivity consideration, and the planetary properties that can (or cannot) be studied using the method, followed by the prospects with the future observational facilities to be available by 2030.

### 3.1. Astrophysical Characterization

#### 3.1.1. Method and Sensitivity

**Radius.** The radius of a transiting planet is measured primarily from the transit depth. (Strictly speaking, the transit depth tells us the planetary radius relative to the star, thus the stellar radius needs to be well constrained in order to obtain accurate information on the planet's radius.)

**Mass.** Masses of transiting planets have been measured through two methods: the Radial Velocity (RV) method and the transiting timing variation (TTV) method. The RV method, which detects the reflex motion of the star due to the planetary orbital motion through the Doppler shift of the high-resolution stellar spectra, measures the product of the planetary mass and sine of the orbital inclination angle $i$, but because the inclination of transiting planets can be determined ($i \sim 90°$), the true mass is obtained. A difficulty with potentially habitable planets is that the variation of stellar RV they produce is generally much smaller than most of the successful RV observations to date. The amplitude of stellar RV variation due to a planet, $K$, is approximately:

$$K \sim 9 \ cm/sec \ \left(\frac{M_p \sin i}{M_\oplus}\right)\left(\frac{a}{1 \ AU}\right)^{-1/2}\left(\frac{1}{\sqrt{1-e^2}}\right) \ \left(\frac{M_\star}{M_\odot}\right)^{-1/2} \qquad [1]$$

where $M_\star$ and $M_p$ are the stellar and the planetary masses, $M_\odot$ and $M_\oplus$ are the solar and the Earth's masses, $a$ is the semi-major axis, and $e$ is the eccentricity. The RV variation of G-type stars orbited by an Earth twin, of the order of 10 cm/s, is challenged by the 'jitter' of the stellar RV originating from magnetic activity, convection, etc. (e.g., Saar et al. 1998, Queloz et al. 2001, Dumusque et al. 2011a, 2011b, Boisse et al. 2012) as well as instrumental noise (see Fischer et al. 2016 for a recent overview of the field). However, the RV amplitude of a late-type star orbited by a HZ Earth-sized planet is larger due to the small stellar mass ($M_\star \leq 0.5 \ M_\odot$) and the small orbital distance of the HZs ($a \leq 0.3$ AU). Such planets have already been discovered through the RV method (Anglada-Escudé et al. 2012, 2016), and the ongoing development of high-resolution spectroscopic instruments will further enforce the discoveries of similar targets (Section 3.1.2). Near-future transit survey by *TESS* will focus on those around nearby late-type stars, providing a synergy with RV observations. While temperate Earth-sized planets around solar-type stars typically suffer from the observational noise, those around bright and quiet stars and/or those on the larger end may become observable. In addition, RV observations will be facilitated once the planetary signal is detected by other means (for example, transit signals obtained with *PLATO*; Section 3.1.2) because the priors for the orbital period and ephemeris will ensure the efficient use of telescopes as well as make the data reduction easier.

If the target planet is in a multi-planet system and the planetary companion(s) is/are observed to transit, planetary mass can also be constrained from the transit-timing variations (TTV) as a result of a mass-dependent dynamical perturbation of the planet (Holman & Murray 2005; Agol et al. 2005). This method is based on photometric data with high time resolution, and is useful when the host star's RV modulation is difficult to observe. Indeed the masses of some of the recently discovered Earth-sized planets have been constrained through this method (e.g., Gillon et al. 2017; Lissauer et al. 2011). TTV may be used to constrain the masses of temperate Earth-sized transiting planets around G-type stars as targeted by *PLATO* where RV methods could be confounded.

**Interior composition.** Once both radius and mass are measured, internal structure may be constrained. The mass-radius relationship allows us to distinguish rocky/metallic terrestrial planets with thin atmospheres, from planets with thick atmospheric envelopes, and perhaps those composed mainly of water (e.g., Léger et al. 2004), although the intermediate densities are confronted by degeneracies (e.g., Fortney et al. 2007; Seager et al. 2007; Rogers & Seager 2010). Combined with stellar composition, inferences on the compositions of rocky material could also be made (Dorn, Hinkel, et al. 2017; Dorn, Venturini, et al. 2017).

*Orbital elements.* The orbital parameters most relevant to the habitability discussion **are** semi-major axis and eccentricity, which determine the incident flux and its time variation. The semi-major axis is constrained from the periodicity of transit light curves provided that the stellar mass is known, while eccentricity will be constrained from RV observations, TTV, and/or occasionally the requirement for stability of the system for multi-planetary systems (e.g., Barnes & Quinn 2001). (The close-in planets are, however, often assumed to be in a circular orbit due to tidal effects.)

Knowing the orbital ephemerides to high precision is essential for the efficient use of telescopes for follow-up observations of known exoplanetary systems as well as for the search of transit signals of RV-detected planets. It has been shown that uncertainties in the eccentricity and argument of periastron can lead to large errors in transit time calculations (Kane & von Braun 2008, Kane et al. 2009). The major issue arises from the uncertainties in the orbital period and time of periastron passage, as well as the time elapsed since the most recent data was acquired, because these cause a drift in phase. In most cases, only a handful of additional RV measurements is needed to provide a dramatic improvement in orbital period and re-sync the location of the planet in its orbit, provided the observations are acquired with the same telescope and instrumentation to remove the need for a data offset.

### 3.1.2. Opportunities through 2030

From space, the re-purposed *Kepler* spacecraft, renamed *K2*, is currently under operation and observing 14 fields near the ecliptic plane in turn to find more transiting planets. *K2* is planned to continue its observations until 2018. There are also ground-based surveys of transiting planets specifically targeting Earth-sized planets around late-type stars, including *MEarth* (Charbonneau et al. 2009; Berta-Thompson et al. 2015) and *TRAPPIST* (Gillon et al. 2016, Gillon et al. 2017).

The primary contributor to the mass measurements of transiting planets is RV observations with ground-based telescopes. In the coming years *a new set of stable high resolution spectrographs* in the visible and near-infrared will be commissioned at 10-m class telescopes and smaller. The *Infrared Doppler instrument* (*IRD*: Y, J, H-bands) for the Subaru Telescope, which adopts laser frequency combs as wavelength calibration to enable extremely high RV precisions, will start operating in 2018 and will be the first ultra-stable spectrograph in the near-infrared range (Tamura et al. 2012). A precursor, the *CARMENES* spectrograph (optical, Y, J, & H-bands), started operating at the Calar Alto 3.5-m telescope in 2016 (Quirrenbach et al. 2016). Two other near-infrared laser-frequency comb-based spectrographs, the *Habitable Planet Finder* (*HPF*: Y & J-bands) for the 9.2m Hobby-Eberly Telescope (Mahadevan et al. 2012) and *SPIRou* (Y, J, H, & K-bands) for 3.6m-Canada-France-Hawaii Telescope (Delfosse et al.

2013), will be ready for use in 2017. Beyond 2018, high-precision spectrographs for planet surveys will be commissioned: *CRIRES* plus for VLT (Follert et al. 2014; Dorn et al. 2016), *iLocater* for the Large Binocular Telescope (LBT; Crepp et al. 2016), *Near InfraRed Planet Searcher* (*NIRPS*) as a near-infrared version of *HARPS* for the 3.6 m-telescope at La Silla Observatory, and the *Echelle SPectrograph for Rocky Exoplanet and Stable Spectroscopic Observations* (*ESPRESSO*) with the VLT, which is the first spectrograph designed with the goal of reaching 20 cm/s for its overall radial velocity precision (Pepe et al. 2014). Eventually, the measurements made by such instruments will be limited by noise imparted by our own atmosphere. If this proves a limiting factor in the detection of Earth-sized planets around Sun-sized stars, then such detections will have to be made from space.

In 2018, *TESS* and *CHEOPS* will be launched to discover nearby transiting planets, with the primary targets being short (< 30 day) orbit planets, including HZ planets around late-type stars.

**The Transiting Exoplanet Survey Satellite (TESS)** (Ricker et al. 2014) is an all-sky, two year Explorer-class planet finder mission launched in 2018, designed to identify planets ranging from Earth-sized planets to gas giants, covering a wide range of stellar types and orbital distances. The main goal of the *TESS* mission is to detect (small) planets around bright host stars that will be good targets for atmospheric characterization with e.g. *JWST*. *TESS* will tile the sky with 26 observation sectors, spending at least 27 days staring at each $24° \times 96°$ sector and observing 200,000 stars, as defined in the *TESS* Input Catalog (TIC). The sectors will overlap at the ecliptic poles, covering the *JWST* Continuous Viewing Zone (CVZ), in order to search for smaller and longer period planets. It was shown that TESS will find approximately 1700 transiting planets from its 200,000 pre-selected target stars—based on simulations of the nearby population of stars, occurrence rates of planets from the *Kepler* mission, models of photometric performance and sky coverage of the *TESS* cameras (Sullivan et al. 2015). Sullivan et al. (2015) also predicted that *TESS* will detect approximately 48 planets with $R_p < 2R_\oplus$ and $0.2 < S_p/S_\oplus < 2$ ($R_p$ and $S_p$ are the radius and the incident flux of the planet, respectively, and $R_\oplus$ and $S_\oplus$ denote Earth's values), around late-type stars with effective temperature lower than 4000 K. Between 2 and 7 of these planets will have host stars brighter than K-band magnitude of 9 and will be very interesting targets for *JWST* to follow up by spectrophotometrically characterizing their atmospheres and searching for the first signs of habitability (Sections 3.2, 3.3).

**The CHaracterising ExOPlanet Satellite (CHEOPS) mission** (ESA/SRE(2013)7; Beck et al. 2016) is the first ESA Small (S-class) mission to perform ultra-high precision photometry of exoplanetary systems. Its main objective will be to search for transits around bright stars known to harbor planets detected via RV measurements. It will aim to determine both whether the known planets transit or not and the transit detection of additional close-in planets not detected by radial velocity. This search will focus on shallow transits on bright stars (6 < V < 9 mag, where V is the V-band magnitude) in the mass range smaller than Neptune with orbital periods of up to ~50 days. When a transit is found, it provides the unique capability of determining radii and therefore densities with ~10% accuracy for these targets. Using the density provided by *CHEOPS*, one can infer the atmospheric volume in a wide parameter space of environmental conditions. *CHEOPS* will also provide improved radii for already known planets and planets that will be discovered by the future space-based or ground-based transit surveys. This sample of well characterized small transiting exoplanets around bright host stars will be a group of targets very well suited for upcoming space-based and ground-based platforms, which focus on spectroscopic characterization of exoplanetary atmospheres.

In the 2020s, transiting planets in a broader parameter space will be surveyed by ***PLAnetary Transits and Oscillations of stars (PLATO)*** (Rauer et al. 2014). *PLATO* has been selected for the M3 launch opportunity (currently planned for 2026) in ESA's Cosmic Vision 2015-2025 program. *PLATO*'s main science goal is to photometrically detect planetary transits and to characterize exoplanets and their host stars, including terrestrial planets in the HZs of solar-type stars, by monitoring up to one million stars covering up to 50% of the sky. Extensive end-to-end simulations have shown that *PLATO* will be able to detect solar system analogs: the discovery of Venus and Earth analogs transiting G-type stars like our Sun is feasible (Hippke & Angerhausen 2015; ESA-SCI(2017)1). Characterization includes the following goals for the uncertainties: 3% for planetary radii, 10% for planetary masses (through radial velocity measurements and TTVs) and 10% for planetary system ages (via asteroseismology of host stars), for planets orbiting bright stars. The resulting large sample of accurately characterized terrestrial planets at orbital periods beyond 3 months will be a unique contribution of *PLATO* to exoplanet research and allow for comparative exo-planetology up to 1 AU orbital distance. Planets orbiting the brightest stars will be key targets for transit spectroscopy of their atmospheres with telescopes such as *JWST* or the ELTs. Radius measurements of individual planets as well as the statistical mass-radius relationship of terrestrial planets in a larger orbital separation found by *PLATO* also serve as a basis in characterizing such targets with future direct-imaging missions, where planetary size is difficult to measure directly. In addition to these exoplanet studies, the large data set of stellar light curves obtained by *PLATO* will allow us to study the stellar structure, evolution, and activity through asteroseismology and rotational modulations, which provide additional science returns into stellar, galactic, and extragalactic research.

## 3.2 Chemical/Climatological Characterization: Transmission Spectroscopy

### 3.2.1 Method and Sensitivity

Transmission spectroscopy is a technique to detect the difference between out-of-transit and in-transit spectra, which can reveal the absorption and scattering properties of planetary atmospheres (Seager and Sasselov 2000). Figure 3 is the transmission spectrum of the Earth observed using lunar eclipse (Pallé et al. 2009), exhibiting the major absorption features of $H_2O$, $O_2$, $O_3$, $CO_2$, and $CH_4$, imposed on a slope due to Rayleigh scattering.

The strength of the spectral features relative to the total stellar flux, is estimated by

$$S \sim \frac{2N_H H R_p}{R_\star^2} \sim 84 \; ppm \; \left(\frac{N_H}{4}\right)\left(\frac{H}{8 \; km}\right)\left(\frac{R_p}{R_\oplus}\right)\left(\frac{R_\star}{0.1 R_\odot}\right)^{-2} \qquad [2]$$

where

$$H = \frac{\mathcal{R}T}{\mu_{atm} g} \sim 7.6 \; km \; \left(\frac{T}{250 \; K}\right)\left(\frac{R_p}{R_\oplus}\right)^2 \left(\frac{M_p}{M_\oplus}\right)^{-1}\left(\frac{\mu_{atm}}{28 \; g/mol}\right)^{-1} \qquad [3]$$

while $R_\star$ and $R_p$ are the stellar and planetary radii, respectively, $\mathcal{R}$ is the gas constant, $T$ is atmospheric temperature, $\mu_{atm}$ is the mean molecular mass of the atmosphere, and $g$ is the surface gravity of the planet. Here, the depth of spectral features is represented by $N_H H$ where $H$ is the scale height of the atmosphere and $N_H$ is a coefficient, which is typically 1-5 for spectral features in the optical to far-infrared range with spectral resolution $\mathbb{R} = 100$-$1000$, depending on the atmospheric composition (e.g., Kaltenegger & Traub 2009). In Eq. [2], we normalized the signal for an Earth-sized planet with an Earth-like $N_2$-dominated atmosphere around a late M-type star with $R_\star \sim 0.1 \; R_\odot$, similar to TRAPPIST-1 (Gillon et al. 2016). If the host star has the Solar radius and other things are equal, the signal would be less than 1 ppm, too small to be detectable in the near future as described below. However, planetary parameters including atmospheric mean molecular mass and the surface gravity vary signal levels.

The detectability of the features depends on the observational strategies, instruments used and the analysis processes. In an idealized case where one tries to identify a spectral feature in a continuum whose only noise source is the photon noise, the signal-to-noise ratio (SNR) is determined by the stellar photon counts $N_\star$ and the signal level $S$ :

$$SNR \sim \frac{N_\star S}{\sqrt{2N_\star}} \sim \frac{S}{\sqrt{2}}\sqrt{\frac{\pi R_\star^2 \dot{n}(\lambda; T_\star)}{d^2} \; \pi \left(\frac{D}{2}\right)^2 \; \Delta\lambda \; \Delta t \; \xi}$$

$$\sim 10 \left(\frac{N_H}{4}\right)\left(\frac{H}{8 \; km}\right)\left(\frac{R_p}{R_\oplus}\right)\left(\frac{R_\star}{0.1 R_\odot}\right)^{-1}\left(\frac{\dot{n}(\lambda; T_\star)}{\dot{n}(3 \; \mu m; \; 2500 K)}\right)^{1/2}\left(\frac{d}{10 \; pc}\right)^{-1}$$

$$\times \left(\frac{D}{6.5 \; m}\right)\left(\frac{\Delta\lambda}{0.1 \; \mu m}\right)^{1/2}\left(\frac{\Delta t}{30 \; hr}\right)^{1/2}\left(\frac{\xi}{0.4}\right)^{1/2} \qquad [4]$$

$$\dot{n}(\lambda; T) \equiv \frac{B(\lambda; T)}{(hc/\lambda)} = \frac{2c}{\lambda^4} \; \frac{1}{\exp\left(\frac{hc}{\lambda k_B T}\right) - 1} \qquad [5]$$

where $\lambda$ is the wavelength, $T_\star$ is the stellar effective temperature, $B(\lambda; T_\star)$ is the black body radiance as an approximation for the stellar spectrum, $\dot{n}(\lambda; T_\star)$ represents the corresponding photon count, $d$ is the distance from the star, $D$ is the aperture of the telescope, $\Delta\lambda$ is the wavelength resolution, $\Delta t$ is the integration time through the transits, and $\xi$ is the total throughput. The factor $\sqrt{2}$ comes from the assumption that the in-transit spectrum is calibrated by out-of-transit spectrum with equal integration time; thus, the observation would require $\sim 2\Delta t$ in total. Again, we adopted stellar parameters similar to TRAPPIST-1, and consider $JWST$ as an example telescope assuming $D = 6.5 m$ and $\xi = 0.4$ (Cowan et al. 2015). Even when considering this idealized situation with planets around late M-type stars, it is likely necessary to accumulate tens to hundreds of hours in total integration time, or tens of transits, in order to detect atmospheric signatures. Such observations can be demanding, and it is therefore critical to have a handful of golden targets that orbit bright host stars and are hence best suited for follow-up observations.

In reality, there exists additional systematic noise that can be instrumental and/or astrophysical (Barstow et al. 2016; Greene et al. 2016). Currently, *HST* and *Spitzer* observations leave tens of ppm as a noise floor that is not reduced after co-adding the data. Given that the expected signal level of atmospheric features can be of the order of 10 ppm or less, the detection of these features may be critically challenged by such noise. Signatures of Earth-twins around solar-type stars are therefore much less likely to be detected.

While we have assumed low-resolution spectroscopy or multi-band photometry above, the past few years have seen fast development in the technique to use high-resolution spectroscopy for characterization of exoplanetary atmospheres. When the resolution is sufficiently high ($\mathbb{R} \gtrsim 100,000$), numerous lines are resolved and the cross-correlation analysis with the modeled template spectra can be performed (see Figure 4 for the example of 1.27 μm $O_2$ features with varying spectral resolutions). The high-resolution transmission spectroscopy has been successfully performed for the Jupiter-size close-in planet HD 209458b by Snellen et al. (2010) using CO features. This technique could be applied to characterizing the atmospheres of Earth-sized planets (Snellen et al. 2013, Rodler and López-Morales 2014). Such high-resolution transmission spectroscopy will be a specialty of ground-based telescopes in the coming decade because none of the planned space-based missions can perform high-resolution spectroscopy. Future 30-meter class telescopes will offer powerful facilities suitable for this kind of observation (Section 3.2.3).

*3.2.2. What can be studied?*

**Gases.** Transmission spectra are sensitive to the constituents of the upper atmospheres (i.e., at low pressures). The signal-to-noise ratio favors the wavelengths where the stellar flux peaks, but may be observable out to the mid-IR range from space, depending on the instrumental sensitivity, the brightness of the star, and the spectral resolution needed. Major molecular features in this range include those from $CO_2$ (2.7, 4.3, 15 μm), $H_2O$ (0.94, 1.13, 1.4, 1.9, 2.7, 6 μm), $O_2$ (0.69, 0.76, 1.27 μm), and $O_3$ (0.5-0.7, 3.3, 4.7, 9.6 μm) (e.g., Kaltenegger & Traub 2009; Bétrémieux and Kaltenegger 2014; Misra et al. 2014b), while Rayleigh scattering produces the characteristic slope at the short wavelengths. Absorption bands of other molecules, which could potentially be important for other worlds, include $CH_4$ (2.3, 3.3, 7.7 μm), CO (2.35, 4.6μm), $SO_2$ (4.0, 7.3, 8.6, 18 μm), $N_2O$ (2.9, 3.9, 4.5, 7.7, 17 μm), $NH_3$ (1.5, 2, 2.3, 3, 6.1, 10 μm), $O_2$-$O_2$ dimer features (1.06, 1.27μm), $CH_3Cl$ (3.4, 7, 10, 14 μm), and DMS (3.4, 6.9, 7.6, 9.7, 14.5 μm); see the upper panel of Figure 2, and Catling et al. (2017) for a more comprehensive list. Also, many organic species have absorption bands in the mid-IR, which would help a search for a broader list of potential biogenic molecules (Seager, et al., 2016).

The diversity of the atmospheric properties of Earth-sized planets is probably large, and the signal levels will likely vary. In particular, as indicated by Eqs. [2] [3], planets with hydrogen-rich atmospheres ($\mu_{atm} \sim 2$) should have increased scale heights and hence have amplified signals overall (see, e.g., Miller-Ricci et al. 2009; Seager et al. 2013; Pierrehumbert and Gaidos 2011; Ramirez and Kaltenegger 2017). Likewise, surface gravity of the planet affects the overall signal strength through the scale height. Geological activities modify the abundance of the molecules involved in the cycling, such as $CO_2$ and $SO_2$ (Kaltenegger et al. 2010a, 2010b, 2013). Interaction with incoming radiation also matters; for example, in an Earth-like atmosphere, $CH_4$ would accumulate more easily under the UV irradiation of an M-type star than a G-type star, improving the expected signal level of $CH_4$ bands (e.g., Segura et al. 2005; Rauer et al. 2011; Hedelt et al. 2013; Rugheimer et al. 2013; Rugheimer et al. 2015). The mixing ratio of $H_2O$ also depends on photochemistry (Rauer et al. 2011; Rugheimer et al. 2013; Rugheimer et al. 2015) as well as the effects of 3-dimensional atmospheric structures (Kopparapu et al. 2017, Fujii et al. 2017a).

In the inverse problem for determining quantitative estimates of molecular abundances, major molecular absorption depths in transmission spectroscopy constrain the relative abundance of the spectrally active molecules, while the mixing ratios of these species and the spectrally inactive components requires the Rayleigh slope unless the higher-order spectral features of absorption bands are measured (Benneke & Seager 2012; Heng and Kitzmann 2017).

Transmission spectroscopy in the UV potentially provides a valuable opportunity to probe the extended exospheres of terrestrial planets. When the planetary atomic exosphere is extended, it can absorb a substantial fraction of the stellar emission lines of the same atom during the planetary transit. Indeed, the absorption signatures in the stellar Lyman-alpha emission line (1215.67Å) due to the extended atomic hydrogen tail of the planetary exosphere have been detected for the warm Neptune-mass exoplanet GJ 436b (Ehrenreich et al. 2015; Bourrier et al. 2016), and it may not be a huge leap to observe smaller, cooler planets once good targets are found. A planet with an ocean that evolves to have a significantly moist upper atmosphere due to, e.g., the increased intensity of the host star, would lead to efficient hydrogen escape to space, potentially resulting in the similar features (Jura 2004). In addition, UV transmission spectroscopy may also be used to study atmospheric molecules, including biosignature candidates such as $O_2$, $O_3$, $H_2O$, $N_2O$, $CH_4$, whose cross sections are significantly larger in UV than in the visible/near-infrared range (Bétrémieux & Kaltenegger 2013). Such an advantage in the atmospheric molecular signatures is, however, at least partially offset by the generally lower stellar flux available in the UV than in the visible and near-infrared range.

***Clouds/haze.*** Cloud/haze layers may be inferred from a broad slope in transmission spectra (e.g., Robinson et al. 2014b), or from muted spectral features (e.g., Kreidberg et al. 2014). Because of the tangential optical paths of transit geometry, even tenuous clouds/haze can contribute to a considerable optical depth and, if present at low pressure, the molecular features can be significantly weakened. While they can be inconvenient obstacles to detections of molecular features, these may also be seen as a signal that could provide insights into the atmospheric compositions (e.g., Hu et al. 2013; Checlair et al. 2016) and have even been proposed as potential biosignatures in certain atmospheric contexts (Arney et al. 2016). At longer wavelengths, transmission spectra are less sensitive to high-altitude haze particles due to the reduced extinction efficiency (e.g., Hu et al. 2013; Arney et al. 2017).

*Vertical structure.* When the stellar light is transmitted through the planetary atmosphere, it is refracted as a result of the atmospheric density gradient. The refraction has notable effects on transmission spectroscopy (Garcia Munoz et al. 2011, 2012a, 2012b). Due to refraction, the altitude at which the transmitted (and refracted) ray probes the atmosphere changes over time, and there is a lower limit in altitude (an upper limit in pressure) to which the transmission spectra are sensitive (García Muñoz and Mills, 2012; Garcîa Muñoz et al., 2012, Bétrémieux and Kaltenegger, 2014, Misra et al. 2014b). Thus, time-resolved tranit spectroscopy, while extremely challenging, would in principle probe atmospheric properties at different altitudes (Misra et al., 2014b). In particular, slightly before or after transit, some fraction of the stellar light refracted through the relatively lower part of the planetary atmosphere reaches the observer if the atmosphere is optically thin, producing an increase in the stellar flux. Such an increase may be used to identify an optically thin atmosphere down to the low altitudes, which is favorable for follow-up observations to detect atmospheric molecules of terrestrial planets (Misra et al. 2014b).

*Surface pressure, temperature.* Transmission spectra could probe the surface pressure if the atmosphere is so thin that atmospheric refraction does not limit our ability to probe the surface layers, and radiatively transparent along the slant path at some wavelengths. However, both factors are likely to prevent us from probing the lower atmosphere of Earth-like atmospheres (Garcia Munoz et al. 2012a; Misra et al. 2014b; Bétrémieux & Kaltenegger 2014), thus the surface pressure is likely to remain unconstrained. The surface temperature would also likely remain unconstrained, while temperature in the upper atmosphere affects the scale height (Eq. [3]).

### 3.2.3. Opportunities through 2030

So far, *HST* has been the most powerful observatory for transmission spectroscopy. *HST* will likely remain the only one capable of observing transmission spectroscopy in the UV in the coming years. A new space UV observatory from Russia, **WSO-UV project**, is planned with a 1.7-meter telescope (Sachkov et al. 2014), which can provide deep transit observation capability in the UV.

Transmission spectroscopy in the visible and near-IR has also been performed with ground-based telescopes with varying spectral resolutions. The new stable visible and near-infrared high-resolution spectrographs with the 10-meter class telescopes (Section 3.1.2) will be a powerful tool to characterize planetary atmospheres using high-resolution transmission spectroscopy, allowing the search for molecular signatures of hot Jupiters and potentially down to Neptune-sized planets.

In 2020, a new space observatory, the **James Webb Space Telescope** (**JWST**), will be launched. *JWST* is NASA's multi-purpose space observatory with a 6.5 meter mirror. One of its main capabilities will be its ability to study the atmospheres of exoplanets with observations in transit, eclipse, or throughout their orbits as a continuous time series to create phase curves. Its halo orbit around the Earth-Sun L2 point allows for long, highly stable, uninterrupted observing sequences compared with ground-based observatories or the *HST*. *JWST* has four instruments: the Near-Infrared Camera (NIRCam), Near-Infrared Spectrograph (NIRSpec), Near Infrared Imager and Slitless Spectrograph (NIRISS), and Mid-Infrared Instrument (MIRI) over its wavelength range of 0.6 to 28 μm at spectral resolution $\mathbb{R} = 4 - 3250$. In principle, all of these instruments can be used to study transiting exoplanets and will provide a spectrophotometric precision of 10-100 ppm for time series observations spanning from hours to days.

Several studies explored the potential for *JWST* to observe the targets provided by *TESS*, *CHEOPS* and other surveys by the time of its launch (e.g. Deming et al. 2009; Batalha et al. 2015; Greene et al. 2016), and the transit community already defined an Early Release Science (ERS) case that focuses on testing relevant observing modes to provide the data and expertise to plan the most efficient transiting exoplanet spectrophotometry characterization programs in later cycles (Stevenson et al. 2016). Following these studies it is anticipated that *JWST* will enable a survey of ~100 gas and ice giants and ~10s of sub-Neptune-sized planets covering a broad range of spectral types, metallicity, and orbital parameters. These results will advance our understanding of the formation and evolution of these planets (see Section 5.3), as well as the nature of possible high-altitude haze/clouds that suppress the molecular signatures.

*JWS*T will also provide the very first opportunity to characterize the atmospheres of temperate terrestrial planets via transmission spectroscopy, spectroscopy of thermal emission (Section 3.3 below), and the orbital phase curves (Section 4.2 below), but only after co-adding tens of transits, or a tens to hundreds of hours in total integration time, depending on the details of the target system (Eq. [4]). For a few nearby systems with late-type stars, first investigations of signs of habitability and isolated, inconclusive biomarkers may be possible if the systematic noise of JWST turns out to be sufficiently smaller than the signal levels. For example, Barstow & Irwin (2016) suggested that an Earth-like ozone layer, if it exists, could be detected in 30 transits by JWST for TRAPPIST-1c and 1d, assuming an Earth-like atmosphere. As an exotic possibility, planets in the HZs of white dwarfs could constitute other golden targets, having significantly larger signals in transmission spectra; for such targets, even the weaker signature of $O_2$ could be observable after a timeframe as short as several hours of integration (Loeb & Maoz 2013). For the prospects of eclipse spectroscopy and phase curves of potentially habitable planets, see Section 3.3 and Section 4.2 below, respectively.

In the 2020s, three **30/40-meter class ground telescopes**, often called **Extremely Large Telescopes (ELTs)**, are planned to operate. These are the *Giant Magellan Telescope* (*GMT*; 24.5 meter diameter), the *Thirty-Meter Telescope* (*TMT*; 30 meter diameter), and the *European-Extremely Large Telescope* (*E-ELT*; 39 meter diameter; now renamed as ELT), expected to be operational in 2023, 2024, and 2027, respectively. A high-resolution ($\mathbb{R} = 25,000 - 120,000$) spectrograph, G-CLEF (Szentgyorgyi et al. 2012), will be installed as one of the first-light instruments on *GMT*, while other telescopes also contemplate ultra-stable high-resolution spectrographs in their instrumentation plans. The prospects for the detection of molecular features in transmission spectra with such instruments depend on the assumptions of the technical specifications for the proposed instruments and telescopes, and the type of noise sources considered. The 0.76 μm oxygen feature of a planet around a nearby (~5pc) late M-type star could be detected after about 100 transits and 30-50 transits with a G-CLEF-like instrument onboard GMT and ELT, respectively, supposing the planet possesses an Earth-like atmosphere (Snellen et al, 2013; Rodler & López-Morales 2014). The nominal specification for the proposed high resolution instrument (HIRES) for ELT suggested that for TRAPPIST-1 b & c, one would be able to detect the 1.3-1.7 μm $H_2O$ band at a SNR of 6 in 2 transits and the 0.9-1.1 μm $H_2O$ band in 4 transits (HIRES team, priv. comm.).

The larger apertures of ELTs will improve photometric precision and hence in principle benefit the conventional low-resolution transmission spectroscopy (Pallé et al. 2011). However, this technique relies on the simultaneous observations of nearby bright stars to correct for atmospheric effects and variability, and the small field of view of ELTs will make it trickier to find suitable comparison stars.

## 3.3 Chemical/Climatological Characterization: Eclipse Spectroscopy

### 3.3.1. Method and Sensitivity

Dayside emission of transiting planets may be identified using secondary eclipses (planet occultation by the star) by taking the difference between the out-of-eclipse and in-eclipse spectra. Figure 5 shows the simulated thermal emission spectrum of the Earth relative to the solar spectrum (black), together with theoretrical spectra of Earth-sized planets with Earth-like atmospheres around different spectral types of stars, modeled with 1D photochemical models (Rauer et al. 2011). As indicated in the figure, the contrast between the planetary flux and the stellar flux in the planetary thermal range, $C_{MIR}$, becomes on the order of 1–100 ppm at $\gtrsim 8$ μm for temperate Earth-sized planets around late-type stars. In this range, features of $H_2O$ (5-8 μm), $CH_4$ (7.7μm), $O_3$ (9.6μm), and $CO_2$ (15μm) are seen. For those around G-type stars, $C_{MIR} \lesssim 1\ ppm$, easily overwhelmed by the expected noise floor. An estimate of the contrast in the thermal regime is given by:

$$C_{MIR}\ (\lambda) \sim 54\ ppm\ \left(\frac{R_p}{R_\oplus}\right)^2 \left(\frac{B(\lambda; T_p)}{B(10\ \mu m;\ 300K)}\right) \left(\frac{R_\star}{0.1R_\odot}\right)^{-2} \left(\frac{B(\lambda; T_\star)}{B(10\ \mu m;\ 2500K)}\right)^{-1} \quad [6]$$

Assuming an idealized photon-noise limited situation with low-resolution spectroscopy, the SNR can be expressed as:

$$SNR \sim \frac{N_\star C_{MIR} \delta}{\sqrt{2 N_\star}} \sim 1.6 \, \delta \left( \frac{R_p}{R_\oplus} \right)^2 \left( \frac{\dot{n}(\lambda; \, T_p)}{\dot{n}(10 \, \mu m; \, 300K)} \right)$$

$$\times \left( \frac{R_\star}{0.1 R_\odot} \right)^{-1} \left( \frac{\dot{n}(\lambda; T_\star)}{\dot{n}(10 \, \mu m; \, 2500 \, K)} \right)^{-1/2} \left( \frac{d}{10 \, pc} \right)^{-1}$$

$$\times \left( \frac{D}{6.5 \, m} \right) \left( \frac{\Delta \lambda}{0.1 \, \mu m} \right)^{1/2} \left( \frac{\Delta t}{30 \, hr} \right)^{1/2} \left( \frac{\xi}{0.4} \right)^{1/2} \qquad [7]$$

where $\delta$ is the relative depth of the spectral features. Again, the fiducial values for the parameters mimic TRAPPIST-1, a late M-type star. We may consider lower wavelength resolution (i.e., larger $\Delta \lambda$) as the spectral features are typically broader in the mid-IR. Still, depending on the configuration, eclipse spectroscopy is as demanding as transmission spectroscopy. Eclipse spectroscopy will not be feasible for planets around solar-type stars nor in the visible/near-infrared range, where the contrast between the star and the planet is smaller than 1 ppm (Eq. [6] below).

### 3.3.2. What can be studied?

**Gases and thermal profile.** Eclipse spectroscopy works best around 8-30 μm where the planet-to-star contrast is large while the planets are not too faint. Signatures of major small molecules in this range include $O_3$ (8.9, 9.6, 14μm), $CO_2$ (15μm), $CH_4$ (7.7μm), $SO_2$ (8.6, 18μm) and $N_2O$ (7.7, 8.6, 17μm); see Figure 2. In addition, volatile organic compounds produced through biological processes also have absorption bands (e.g., Domagal-Goldman et al. 2011) . Compared with transmission spectrum, emergent thermal emission can probe deeper atmosphere due to the short optical path length and is less likely to be obstructed by tenuous haze layers.

The molecular features in thermal emission depend not only on the abundance of molecules but also upon the temperature profile of the atmosphere. The decreasing temperature as a function of altitude results in absorption features, while thermal inversion layers can create emission features. If the planet does not have a strong vertical temperature gradient, molecular features are weakened. Thus, some information of vertical temperature gradient can be obtained. Once the detailed features of line shapes could be resolved, which is unlikely through 2030, they would further constrain the vertical temperature profiles. If the atmosphere is optically thin at some thermal wavelengths, thermal emission spectra can constrain the surface temperature, one of the key factors for habitability.

While thermal emission spectra are sensitive to these properties, the interpretation as an inverse problem may not be straightforward (for a retrieval study for a cloud-free Earth-like atmosphere, see von Paris et al. 2013). The presence of clouds and the 3-dimensional heterogeneity further complicates the problem. The full retrieval would require sophisticated parametric models of atmospheres as well as high-precision observations to feed into the models.

The ingress and egress light curves of the planetary eclipse have offered the opportunity to obtain 2D maps of the dayside of hot Jupiters (Majeau et al. 2012, de Wit et al. 2012). Applying to Earth-size planets, however, would be exceedingly difficult due to the weakness of the planetary signal.

**Solid surface.** If the atmosphere is optically thin, we will see the spectroscopic features of the surface rocky materials. Notable features in the mid-infrared include bands from Si-O bonds of rocky materials around 10 μm and 20 μm (Hu et al. 2012a; Figure 2).

### 3.3.3. Opportunities through 2030

Since the eclipse spectroscopy of potentially habitable planets favors the mid-infrared observations, space observatories work best. So far, most eclipse spectrophotometry of Jupiter-like planets has been performed with the *Spitzer* space telescope. In the near future, ***JWST*** (Section 3.2.3) will be the most promising observatory. However, a smaller-than-predicted noise floor would be necessary to detect spectral features in the thermal emission through eclipse observations.

## 4. Characterizing Planets with General Orbital Inclination

In this section, the methods to characterize potentially habitable planets with general orbital inclination are considered. While non-transiting planets are missed by the observation techniques unique to transiting ones, they are in general closer to the Earth, benefiting other follow-up observations. In the following, astrophysical characterization of non-transiting planets (Section 4.1), chemical, climatological characterization through phase curve measurement (Section 4.2), high-contrast imaging (Section 4.3), and the spectral method (Section 4.4) are discussed. The format is similar to Section 3.

## 4.1. Astrophysical Characterization

**Radius.** Although radius is one of the very basic properties of planets, radii of non-transiting planets are difficult to measure directly in the foreseeable future. When planets are imaged in the visible to near-infrared range where the scattered light dominates, the disk-integrated intensity is essentially proportional to squared radius times planetary albedo (Eq. [11] below), so the radius is in general degenerate with albedo. For synchronously rotating atmosphere-free planets, planetary radii (as well as albedo) may be estimated from the phase curves (Maurin et al. 2012), although such planets will not be regarded as potentially habitable. Disk-integrated spectra in the mid-infrared (i.e., thermal emission) could better constrain planetary radius (e.g. Des Marais et al. 2002), but currently there are no projects that are capable of such observations (Section 6.2.1).

**Mass.** Planetary mass can be estimated through the RV method, as in the case of transiting planets. Without another type of observation, the degeneracy between the planetary mass and the inclination angle cannot be disentangled (Eq. [1]), while statistically the expected value of the true planetary mass is $4/\pi$ of the measured value for $m_p \sin i$. The true mass can be obtained if the inclination is constrained from other types of observation e.g., multi-epoch direct-imaging observations (Section 4.3) or the Doppler shift of the planetary spectra (Section 4.4). Regarding the sensitivity to HZ Earth-sized planets, the same argument holds as described in Section 3.1: those around solar-type stars are more challenging, while those around nearby late-type stars will probably be accessible.

Another potential probe of planetary mass is astrometry, the method to detect the reflex motion of the star due to the planetary orbital revolution as a periodic movement of the star along the celestial sphere. Astrometry is used by the ongoing *Gaia* mission to discover large, long-period planets, but *Gaia* is unlikely to detect temperate Earth-sized planets (Perryman et al. 2014). The capability of astrometry detecting temperate Earth-sized planets is being discussed in the context of the *LUVOIR*-type far-future mission concept (Section 6.1.2).

**Orbital elements.** In a similar manner to the case of transiting planets, semi-major axis and eccentricity will be constrained from RV observations if detected. Otherwise, multi-epoch direct-imaging observations (Section 4.3 below) can constrain the orbits.

As discussed in Section 3.1, orbital ephemeris is important for follow-up observations. For direct imaging observations, the information of orbital ephemeris enables predictions of the timing of the maximum angular separation, and thus accurate ephemerides from RV data will help optimize the use of the precious hours of high-demand space telescopes (Kane 2013). Orbits of longer-period planets are more difficult to analyze because the uncertainty in the planetary parameters (in particular orbital period and epoch of periastron passage) are large, and data across multiple orbits will substantially improve the estimates.

## 4.2 Chemical/Climatological Characterization: Phase Curves

### 4.2.1. Method and Sensitivity

Planetary spectra that vary as a function of the star-planet-observer angle (*phase angle*) may be extracted as a time-varying component of the star+planet spectra, synchronous to the planetary orbital period. In general, variation amplitude is larger for planets whose orbit is closer to edge-on. Thus, while it can be used for non-transiting planets, transiting planets are the most favorable targets. Phase curve variations would in principle exist both in scattered light and in the thermal emission, but the scattered light of potentially habitable planets is less than 1 ppm of the stellar light and unfeasible to detect. The thermal emission phase curves of those around late-type stars tend to have the best star-to-planet contrast and are most likely to be detectable. Even for such systems, the contrast is on the order of 10-100 ppm, and the phase variation amplitude is smaller than that, so this level of the long-term stability of the stellar radiation and the instruments is a prerequisite for a successful observation. The SNR estimate of thermal emission phase curves in an idealized photon-noise-limited case would be similar to that of secondary eclipse (Eqs. [6] [7]), except for the replacement of $\delta$ by the relative amplitude of phase curves. However, the phase variation itself may be searched for in broadbands (i.e., larger $\Delta\lambda$), which can loosen the observational demands.

### 4.2.2. What can be studied?

**Heat re-distribution--atmosphere/ surface flow.** The broadband thermal emission phase curves are a useful probe of the thermal redistribution across the globe, which may constrain the potential presence of atmospheres (or perhaps a flow on the surface) (Knutson et al. 2007; Demory et al. 2016). For example, atmosphere-less planets exhibit strong day-night contrast in thermal emission, which results in a large phase variation amplitude, while planets with thick atmosphere tend to have horizontally more uniform emission temperatures, minimizing the phase variations (e.g., Selsis et al. 2011). The phase curves also depend on the spin state, thermal inertia, and eccentricity. For example, with non-zero thermal inertia, non-synchronously rotating planets exhibit more modest horizontal temperature gradient, hence smaller phase variations, than synchronously rotating planets (e.g., Selsis et al. 2013).

**Clouds.** Thermal phase curves are affected by also trace the large-scale cloud patterns if they exist. Interestingly, synchronously rotating planets covered with oceans tend to develop thick clouds in the substellar region, and produce characteristic patterns in the orbital phase curves when highly irradiated (Yang et al. 2013; see also Hu and Yang, 2014 for a more realistic treatment of ocean dynamics); this could even indirectly imply the underlying surface liquid water.

**Gases.** Spectrally resolved phase curves, or "variation spectra" (Selsis et al. 2011), imprint the signatures of atmospheric molecules because their wavelength-dependent opacity changes the pressure level at which the phase curves probe, and the different pressure levels may have different horizontal patterns of temperature (e.g., Stevenson et al. 2014). The list of the potential target molecules is similar to that for thermal emission eclipse spectroscopy (see Section 3.3.2).

### 4.2.3. Opportunities through 2030

As in the case of eclipse spectroscopy, ***JWST*** is a promising observatory for thermal emission phase curves of exoplanets, potentially down to temperate Earth-sized planets around nearby late-type stars. The thermal phase variation amplitude of Proxima Centauri b is estimated to be on the order of 10 ppm or less if it possesses an Earth-like atmosphere, and ~100 ppm if it is atmosphere-less, assuming 60 degree inclination (e.g., Turbet et al. 2016, Boutle et al., 2017 Kreidberg & Leob et al. 2016). The large amplitude in the atmosphere-less case would be observable with the Low-Resolution Spectrograph (LRS) mode of MIRI onboard JWST, depending on the noise floor, and the stability of the stellar radiation, and could provide the clues to the potential presence of an atmosphere.

## 4.3. Chemical/Climatological Characterization: High-Contrast Imaging

*4.3.1 Method and Sensitivity*

Direct imaging allows access to exoplanets at all orbital inclinations, thus offering the only path to characterizing the full suite of exoplanets in the solar neighborhood. However, direct imaging of exoplanets separated from the host star is greatly complicated by stellar glare and must rely on instruments to block the on-axis light from target stars while re-directing the effects of diffraction. Coronagraphs and starshades are currently two possible starlight suppression approaches. The former is placed within the payload of a space telescope (i.e., an "internal" occulter) whereas the latter is its own spacecraft positioned many tens of thousands of kilometers away from a space telescope (hence an "external" occulter).

For both methods, the smallest angular separation they can probe for the faint planetary signal (inner working angle or IWA) should be smaller than the largest angular separation between the planet and the star, $\Delta$, which is:

$$\Delta = 100 \ mas \left(\frac{a}{1 \ AU}\right)\left(\frac{d}{10 \ pc}\right)^{-1} \quad\quad [8]$$

where "mas" stands for milliarcsecond. The IWA for coronagraphs is expressed by

$$IWA = 103 \ mas \left(\frac{\mathcal{R}}{2}\right)\left(\frac{\lambda}{0.6 \ \mu m}\right)\left(\frac{D}{2.4 \ m}\right)^{-1} \quad\quad [9]$$

where $\mathcal{R}$ is the minimum number of beamwidths between the star and the planet and is a function of the coronagraph design, and ranges from 2 to 4 at present. For starshades, the IWA is approximately

$$IWA = \frac{2F\lambda}{D'} \sim 73 \ mas \left(\frac{F}{10}\right)\left(\frac{\lambda}{0.6 \ \mu m}\right)\left(\frac{D'}{34 \ m}\right)^{-1} \quad\quad [10]$$

Where $F$ is the dimensionless Fresnel number (about 10 for $10^{-10}$ contrast) and $D'$ is the starshade diameter. These relations mean that, for a fixed-size telescope or starshade, the number of accessible exoplanets at longer wavelengths is greatly reduced relative to the number accessible at shorter wavelengths. Both methods are thus most effective for observations in the visible range where habitable exoplanets are seen in scattered light; the scattered light spectrum of the Earth is shown by the solid line in Figure 6 (Robinson et al., 2011).

Both methods must suppress the stellar light to the level of planetary light. The contrast between the scattered light of the planet (at the maximum separation) and the star, $C_{VIS-NIR}$, is:

$$C_{VIS-NIR}(\lambda) \sim \frac{2}{3\pi} \frac{R_p^2}{a^2} A \sim 10^{-10} \left(\frac{R_p}{R_\oplus}\right)^2 \left(\frac{a}{1\,AU}\right)^{-2} \left(\frac{A(\lambda)}{0.3}\right) \qquad [11]$$

where $A$ is planetary albedo. The contrast for temperate planets around late-type star is improved to $10^{-9}$-$10^{-6}$, corresponding to the smaller orbital distance (0.01-0.3 AU), while the angular separation from the star becomes smaller.

Figure 7 diagrams the contrast and the maximum separation from the host star of the solar system planets at 10 pc and known exoplanets (points), as well as the performance of the existing and future instruments (lines; detailed below). Ongoing efforts are pushing the sensitivity from upper-right corner toward lower-left, where an Earth-twin resides.

A coronagraph camera has a complex optical train that controls diffraction by using one or more image focal and/or pupil planes to block and beam-shape the on-axis starlight with small specialized masks. Coronagraphs have been used for decades in solar observations, have flown on *HST*, and will be flown on *JWST*, and are now used effectively with ground-based telescopes equipped with adaptive optics. The best operational contrast sensitivity achieved to date, from either the ground or from space, is 3 x $10^{-7}$ at 0.4 arcsecond separation by the *SPHERE* instrument on the Very Large Telescope observing Sirius (Figure 7).

In space, a coronagraph with a wavefront sensing and control system, as is now planned for *WFIRST*, can achieve much higher contrasts (neither *HST* nor *JWST* are equipped with this capability). Outside the atmosphere, wavefront correction is much more precise than from the ground, and the system will be stable for hours or days as opposed to the millisecond timescales of atmospheric disturbances. In ground vacuum tests, coronagraphs augmented with wavefront control have already demonstrated 6 x $10^{-10}$ contrast with a 10% bandwidth in a 284 $(\lambda/D)^2$ field extending from 3–15 $\lambda/D$ (Trauger et al. 2013). Work is ongoing to bring the performance to the $1 \times 10^{-10}$ requirement at an IWA of 3 $\lambda/D$ (NASA's Exoplanet Exploration Program Decadal Survey Testbed; N. Siegler, priv comm).

A starshade is an external occulter (Fig. 8), and the effects of diffraction are controlled by the precise analytical shape of its petals and the diffracted light is re-directed away from the receiving telescope, creating a dark shadow for which the telescope can fly in formation. Star light scattered from orbiting planets arrives off-axis, misses the starshade, and can be captured by the telescope (Figure 8); for a more detailed explanation, see the NASA-chartered Exo-S probe study report (Seager et al. 2015).

Starshades present several key advantages over coronagraphs. First, for a given telescope diameter, a starshade can achieve smaller IWAs (Eq. [10]). The smaller IWA enables probing angular regions closer to the targeted stars as well as a larger sample of stellar distances. Second, the achievable contrast ratio is independent of the architecture of the telescope. While segmented and obscured telescope apertures currently pose diffraction challenges for coronagraphs attempting to reach contrast ratios smaller than $10^{-9}$ at visible wavelengths (a topic of active research; NASA's Exoplanet Exploration Program Segmented Coronagraph and Design Analysis study; https://exoplanets.nasa.gov/exep/technology/TDEM-awards/), starshades can in theory create $10^{-10}$ or better dark regions at telescope detectors independent of the aperture architecture, reaching the sensitivity requirement to directly image temperate Earth-sized exoplanets around Sun-like stars. Lastly, starshades can be designed to perform at larger bandwidths and, with fewer optics, significantly higher throughput, thus potentially enabling higher resolution spectrographs. A drawback, on the other hand, is that the complex positioning of the two spacecraft and re-pointing will take time—approximately 1-2 weeks—potentially challenging repeat observations for orbit determination and the measurement of seasonal changes and phase angle effects.

The requirement of high-contrast imaging could be loosened when a starlight suppression instrument is combined with spectral separation methods (Section 4.4) (Sparks & Ford 2002; Riaud & Schneider 2007; Kawahara et al. 2014; Snellen et al. 2015). Several works (Kawahara & Hirano 2014; Snellen et al. 2015; Wang et al. 2017) have investigated the potential of the combination of high contrast imaging and high-resolution ($\mathbb{R} \sim 100,000$) spectrograph, and showed that molecular signatures of Earth-like planets around late-type stars may be detected with high-contrast imaging instruments that achieve $\sim 10^{-5}$-$10^{-4}$ contrast (improved by a factor of $10^3$), when combined with high-resolution spectrograph. This is probably a promising approach to spectral signatures of potentially habitable planets with ground-based observatories. See the next section (Section 4.4) for more discussions.

Another way to increase the detectability of the planet is to utilize polarized light, taking advantage of the fact that stars have low polarizations. Ideally, if the polarization from the star is sufficiently low and the precision of the polarimetry is high, polarimetry alone would allow us to identify the planetary component in the combined star + planet light. Indeed, high-precision polarimetric observations attempted to detect scattered light of hot jupiter HD 189733 without starlight suppression (Berdyugina et al., 2008, 2011; Wiktorowicz 2009; Wiktorowicz et al., 2015; Bott et al., 2016); no conclusive detection has been obtained so far, however, while the upper limits of tens of ppm was put. The polarizations of potentially habitable planets are expected to be at even lower level, and polarization on its own will likely be insufficient for planet detection. Instead, a combination of direct-imaging instruments and polarimetric instruments is a more likely way to utilize this technique.

*4.3.2 What can be studied?*

Once the starlight is sufficiently suppressed, and the exoplanet lies outside of the instrument's IWAs, then the exoplanetary photons can be directly analyzed in the spectral and time domains. It can also provide, in a single observation, the full system context view of multiple exoplanets and zones of dusty debris.

*Gases.* Spectroscopy of Earth-like planets in the visible- to near-infrared range show the molecular signatures including $O_2$ (0.63, 0.69, 0.76, 1.27 μm), $O_3$ (< 0.35, 0.5-0.7 μm), $H_2O$ (0.72, 0.82, 0.94, 1.13, 1.4, 1.9 μm), $CO_2$ (1.04, 1.20, 1.43, 1.6, 2.0, 2.7 μm), $CH_4$ (0.72, 0.78, 0.88, 0.97, 1.15, 1.4, 1.66, 2.3 μm) (see Fig. 2), while the features at longer wavelengths reduce the number of accessible targets due to the limitation of IWA. A massive $O_2$-dominated atmosphere would produce strong $O_2$-$O_2$ features (1.06, 1.27μm), potentially diagnostic for a false-positive scenario for $O_2$ as a biosignature (Schwieterman et al., 2016). Direct-imaging spectroscopy is more sensitive to the compositions in the lower atmospheres than transmission spectroscopy, and is less affected by tenuous haze layer or high clouds, due to the shorter optical path length.

The spectral resolution ($\mathbb{R}$) required for a detection clearly varies from band to band. For example, assuming an Earth-like atmosphere, $\mathbb{R} > 20$ will be necessary to detect $H_2O$ absorption bands at 0.94 μm, while the convincing detection of the narrower $O_2$ band at 0.76 μm likely requires $\mathbb{R} > 100$ (Brandt & Spiegel 2014). Some scattering and absorption features are broad enough to be captured with fairly low-resolution spectroscopy or multi-band photometry. For example, Rayleigh scattering, scattering/absorption by clouds and haze layers broadly affect the wavelength dependence of the scattered light spectra (Hu et al. 2013; Arney et al. 2016; Checlair et al. 2016). Broad absorption features of $O_3$ around 0.6 μm may be inferred from the colors of the Earth (Krissansen-Totton et al. 2016).

*Solid surface.* Scattered light spectra include characteristic features of the surface, if the atmosphere is optically thin at least in some areas. Reflectance spectra of commonly seen materials in the Solar System are shown in the lower panel of Figure 2. Spectroscopic features of surface materials include absorption bands of rocky materials due to charge transfer (<0.4 μm) and crystal-field effects (around 1 μm) and of O-H bonds of ice and hydrated materials (1.5 μm, 2 μm, 3 μm), whose exact optical properties depend on the detailed composition and grain size (e.g., Ford et al. 2001; Hu et al. 2012a; Fujii et al. 2014). Importantly, for the first time, we will be able to access surface reflectance biosignatures such as vegetation's red edge (Ford et al. 2001; Seager et al. 2005; Montañés-Rodríguez et al. 2006; Tinetti et al. 2006; Kiang et al. 2007; Kiang et al. 2007), "purple edge" (Sanromá et al. 2014), and the peculiar reflectance spectra that occur in a variety of biological pigments of diverse functions (Hegde et al. 2015; Schwieterman et al. 2015), some of which are shown in Figure 2; see Schwieterman et al. (2017) for a comprehensive review. It would be possible that exo-biospheres interact with incident radiation to imprint peculiar signatures, similar to what we observe on the Earth.

*Surface liquid bodies.* The most essential aspect of climatological characterization for HZ planets is the presence of an ocean. This can be probed in scattered light through the peculiar anisotropy of scattering by liquid surface, where the reflectivity increases with grazing incident angle (ocean's "glint", Williams & Gaidos 2008; Oakley & Cash 2009; Robinson et al. 2010; Robinson et al. 2014a, Robinson 2017). This nature of an ocean exhibits itself as an anomalous increase in planetary albedo at the crescent phase. At such a phase, however, the angular separation between the star and the planet is small and the scattered light is dark, making direct-imaging observations challenging, possibly requiring the direct-imaging missions beyond 2030 (Section 6).

*Planetary albedo.* Because the planetary scattered light is proportional to $R_p^2 A(\lambda)$, the absolute value of planetary albedo $A(\lambda)$ is known in a model-independent manner only when the planetary radius is known from e.g., transit observations, or perhaps from the assumption of interior composition if the mass is known from e.g., RV observations. Planetary albedo yields the equilibrium temperature, an important reference point for modeling of the temperature profile.

*Surface pressure, temperature.* A possible indicator of surface pressure in scattered light spectra is the Rayleigh scattering feature. Rayleigh scattering slope depends on the molecule-specific cross section and the column number density of the atmosphere, and the latter is related to the atmospheric pressure through the surface gravity and the mean molecular weight of the atmosphere. Thus, estimates of these parameters are degenerate. The presence of clouds as well as the wavelength-dependence of surface reflectance can further puzzle the interpretation. The atmospheric pressure also affects the absorption features of molecules. Additional use of signatures of dimers, which vary as the square of the density, may be useful to constrain surface pressure (Misra et al. 2014a). Despite the direct relevance to habitability, surface temperature is difficult to estimate directly from the scattered light spectra, as it negligibly affects the spectra in this regime (Robinson 2017).

*Spin parameters.* Time-resolved direct-imaging observations at the time-scale of planetary rotation, if/once they eventually become feasible, potentially provide additional, key dimensions in climatological characterizations. Unless the surface is completely uniform, rotation rate can be measured as a periodicity of the disk-integrated planetary light; rotation rate is one of the fundamental parameters in modeling the climate and habitability of the planets. The Earth as a point source changes its albedo by 10-20% in one rotation (Livengood et al. 2011, also shown in Figure 9), and Pallé et al. (2008) found that the periodicity can be identified through the auto-correlation analysis despite the variable cloud cover, thanks to the consistent geographic features (e.g., continents). Another important parameter affecting climate, obliquity, could be inferred through examining the long-term light curves which involve rotational and orbital variations (Kawahara and Fujii, 2010, 2011; Fujii and Kawahara, 2012; Kawahara, 2016; Schwartz et al., 2016). The rotation rate of Jupiter-like planets has been measured by the broadening of the molecular lines using high-resolution spectroscopy (Snellen et al. 2014), but similar observations of an Earth-like planet would be exceedingly difficult; the rotation velocity of the Earth is ~26 times smaller than Jupiters.

***Surface heterogeneity, partial cloud cover***. From the rotational variation of scatted light spectra, the heterogeneity of the surface composition may be constrained (e.g., Cowan et al. 2009; Cowan et al. 2011; Cowan & Strait 2013, Fujii et al. 2010; Fujii et al. 2011; Fujii et al. 2017b). For example, combinations of a liquid surface, particulate rocky materials, and/or snow/ice contribute to a marked contrast in scattered light spectra at different wavelengths (Figures 2, 9). The surface heterogeneity may imply geological processes over the history of the planet, including plate tectonics and volcanic activities (Fujii et al. 2014). The presence of variable clouds could be probed through the deviation of daily lightcurves from the average light curve (Pallé et al. 2008), or the variability of molecular features (Fujii et al. 2013). Rotational and orbital variations together would even allow us to recover the 2-dimensional surface map (Kawahara & Fujii 2010, 2011; Fujii & Kawahara 2012).

***Additional clues from polarization.*** Once the planet is directly imaged, the polarization of the planetary light may be analyzed when combined with the proper instrument. Polarimetry of the Earth includes some interesting features. In theory, Rayleigh scattering by the atmosphere causes a polarization peak near a phase angle 90 degrees, whereas the polarization of reflection by a liquid water surface is the largest at phase angle 106 degrees, and water droplets have their polarization peak at around 30-40 degrees, while multiple scattering reduces the polarization (Stam et al. 2004; Stam 2008; McCullough 2006; Bailey 2007; Williams & Gaidos 2008; Zugger et al. 2010; Zugger et al. 2011; Karalidi & Stam 2012). Earthshine observations confirmed that the effect of Rayleigh scattering dominates in the disk-integrated polarized light of Earth at short wavelengths (Sterzik et al. 2012; Takahashi et al. 2013), and the polarization decreases to ~9-12% in the near-infrared continuum (Miles-Páez et al. 2014). The "absorption" features of atmospheric molecules, which exhibit themselves as "peaks" in polarized light because of the reduced contribution from multiple scattering (Stam 2008), are probably as high as 30% (Miles-Páez et al. 2014). These features potentially provide additional evidence about the planetary surface environment.

### 4.3.3 Opportunities through 2030

Direct imaging of Jupiter-sized distant planets has been successfully performed with 10-meter class ground-based telescopes, and these observations achieved a contrast of $10^{-4}$–$10^{-6}$. This level of high-contrast technique could potentially be combined with high-resolution spectrograph to obtain spectral features of smaller planets; we discuss such an approach in Section 4.4. The 30-meter class telescopes (ELTs) to be built in the 2020s are also considering coronagraphic instruments, and depending on the performance of such instruments, they might achieve high-contrast imaging of potentially habitable planets even without high-resolution spectroscopy. Their large apertures can make the inner working angle as small as 10 mas in principle (Eq. [9]), which allows us to probe potentially habitable planets around nearby late-type stars. The accessible targets may also include temperate Earth-size planets around M-type stars in the prolonged pre-main-sequence stage, such as 40 Myr AP Col, 8.4 pc away (Ramirez, Kaltenegger 2014). Once the habitable planets are detected in high-contrast imaging, an Earth-like $O_2$ absorption feature at 1.27 µm would be detectable (Kawahara et al. 2012).

Temperate Earth-sized planets orbiting solar-type stars, however, may never be directly imaged from the ground, due to the $10^{-10}$ contrast requirement, which is limited by Earth's atmosphere and the residual uncorrected wavefront error (Traub & Oppenheimer 2010; p. 146-147). To achieve such contrast, a space-based observatory will be necessary. Although *JWST* is equipped with coronagraphic instruments to perform high-contrast imaging of Jupiter-like planets, they will not be able to image Earth-sized planets.

The next space-based opportunity for high-contrast imaging is ***The Wide Field InfraRed Survey Telescope (WFIRST)***, which is NASA's next large space observatory with an expected launch date of 2025 and will operate for at least six years. Its prime instrument is a 288 megapixel near-infrared camera designed to survey the extragalactic sky for dark energy science, and to monitor the galactic bulge for gravitational microlensing events. In 2013 the mission was redesigned to use a surplus 2.4 m telescope and, with the larger aperture, a second instrument was added: an optical coronagraph for exoplanet direct imaging. The coronagraph instrument ("CGI") will be the first space-based demonstration of precision wavefront control that is needed to achieve an image contrast ratio approaching a billion to one at sub-arcsecond separations. CGI includes both Shaped Pupil and Hybrid Lyot coronagraph masks, a pair of 48x48 actuator deformable mirrors, and a low-order wavefront sensor to compensate for telescope instabilities. CGI will enable studies of dozens of giant planets in scattered light around stars within ~ 20 pc of the Sun, using photometry or $\mathbb{R}\sim50$ spectroscopy over a 0.6-0.95 µm bandpass. The CGI will not be capable of studying Earth twins, but it will have the contrast sensitivity and angular resolution to study larger planets if they are present in the habitable zones of half a dozen nearby stars. With luck, *WFIRST* CGI could obtain the first imaging detection and crude spectroscopy of large temperate terrestrial planets.

While not baselined, a starshade flying in formation with *WFIRST* (Figure 8) offers the critical capability of directly imaging terrestrial planets in the habitable zone of solar-type stars. The 2015 NASA-chartered starshade probe study was extended to optimize the exoplanet yield of a technology demonstrator *WFIRST* Starshade Rendezvous mission. The most likely scenario (included in the extended study's final report (https://exoplanets.nasa.gov/internal_resources/225/)), factoring in observational completeness, predicted 28 habitable zones of solar-type stars could be surveyed by *WFIRST* for Earth-sized exoplanets with a 34 m starshade over a 3-year mission. Assuming a conservative occurrence rate of potentially habitable planets ($\eta_\oplus$) of 10%, stellar sky brightness of 21 mag/sq arcsec (zodi and exo-zodi), and an optical throughput of 28% using the *WFIRST* CGI optics, 2-3 exo-Earth detections were predicted with a SNR of 6. Of those, one could have their spectrum taken within a visible bandpass. Assuming $\mathbb{R}\sim70$, spectral features such as water vapor, oxygen, and potentially even Rayleigh scattering of an exo-Earth atmosphere could be identified. These signatures, along with cloud parameterization and other assumptions, could provide quantitative constraints on oxygen and water vapor mixing ratios along with surface pressure (Feng et al., in prep). Smaller and less expensive "technology demonstrator" starshades (20 m) with shorter mission lifetimes were also included in the extended study resulting in only one predicted imaged terrestrial planet and without characterization. A final specific starshade design will be developed over the next couple of years by NASA starshade technology activity managed by the Exoplanet Exploration Program.

## 4.4. Chemical/Climatological Characterization: Spectral Separation

### 4.4.1. Method and Sensitivity

Spectral methods may also be used to extract the planetary component from the combined star + planet spectra. In high-resolution spectra ($\mathbb{R} \gtrsim 100{,}000$) where individual lines are resolved (Figure 4), the Doppler shifted planetary atmospheric lines with respect to the telluric lines and the stellar lines allows one to identify the planetary component, along with the planetary line-of-sight velocity (Snellen et al. 2010). While the planetary signal is buried deeply in the overwhelmingly brighter stellar flux, cross-correlation analysis of the observed spectra with the modeled ones using many lines contributes to an enhanced signal-to-noise ratio. Such a technique was successfully demonstrated by the molecular detections of hot Jupiters without using planetary transits, eclipses, or high-contrast imaging. For example, Brogi et al. (2012) detected the CO lines of τ Boötis b through high-resolution spectroscopy with the Cryogenic Infrared Echelle Spectrograph (CRIRES) at Very Large Telescope, while de Kok et al. (2013) detected CO lines of HD189733b using the same instruments. The stabilized instruments such as HARPS and HARPS-N have also been used with smaller telescopes (3.5 m) to make the first detection of the scattered light of hot Jupiter 51 Peg b (Martins et al. 2015).

This method will be limited to ground-based observations in a foreseeable future, as high-resolution spectrographs are not planned to be on a space observatory. However, the contrast between temperate Earth-sized planets and their host stars at the wavelengths observable from the ground ($\sim 10^{-10}$-$10^{-6}$ ; Eq. [11]) is much smaller than the successful observations so far ($\sim 10^{-3}$). Thus, several works (Sparks & Ford 2002; Riaud & Schneider 2007; Kawahara et al. 2014; Snellen et al. 2015; Wang et al. 2017) have proposed the combination of the high-resolution technique with a high-contrast imaging instrument as a viable approach for detecting atmospheric signatures of temperate Earth-sized planets around late-type stars. After suppressing the stellar component to the order of $10^{-5}$ or so, spectral identification of the planetary component will be more feasible.

Recently, Snellen et al. (2017) proposed a related technique where space-based medium-resolution spectroscopy is used to find spectral features of the planetary atmosphere in the combined star+planet spectrum in the mid-IR range where the planet-to-star contrast can be order of 10-100 ppm (for those around late-type stars). While the individual lines are not fully resolved in medium-resolution spectroscopy, the high-frequency features may be identifiable by fitting with theoretical models.

One could also attempt to identify the molecular signatures of planetary atmospheres even in the low-resolution star + planet spectra, if the molecules should not be present in the stellar atmosphere and thus can be safely attributed to the planetary origin. The success of this method will critically rely on accurately knowing the host star spectrum as well as on the noise floor of the observed data.

### 4.4.2. What can be studied?

Once the Doppler-shifted planetary lines are detected, not only the presence of targeted molecules is revealed, but also the planetary line-of-sight velocity is obtained. This translates to the true mass of the planet and orbital inclination when combined with stellar RV measurements, contributing the astrophysical characterization of the system.

In the visible to near-infrared range where the scattered light dominates over thermal emission, the contrast between the planet and the star would indicate $R_p^2 A$. In the mid-IR regime where the thermal emission dominates, the detailed characteristics of the spectral features can probe on the vertical thermal profile of the atmosphere, as discussed in Section 3.3;

### 4.4.3 Opportunities through 2030

By combining the ultra-stable high-resolution spectrograph with the high-contrast imaging techniques, the largest existing ground-based telescopes might eventually acquire the potential to access the nearest temperate Earth-sized planets. Lovis et al. (2017) estimated that detection of the scattered light of Proxima Centauri b (37 mas separation from the host star) in the visible wavelengths may be possible with the proposed upgrades to SPHERE high-contrast imager combined with the ESPRESSO high-resolution spectrograph, after 20-40 nights of total telescope time, assuming an Earth-like atmosphere. Even marginal constraints on $O_2$ may be obtained with an intense use of the telescope.

The **ELT**s can in principle adopt the same technique. Their larger collecting areas can substantially improve the observational capabilities. For example, it has been estimated that the atmospheric characterization of Proxima Centauri b could be accomplished in about 6 nights using the collecting area of the ELT (HIRES team, priv. comm.). The larger aperture also allows us to probe a greater number of targets with the smaller IWA. Once coronagraphic instruments and high-resolution spectrographs are installed on the ELTs, they will offer promising and unique opportunities for the targets around late-type stars.

In the mid-IR range where potentially habitable planets around late-type stars have the planet-to-star contrast of the order of 10-100 ppm, spectral signatures of the planetary atmosphere could be searched for in the star + planet spectrum even without starlight suppression instruments, if the observations have sufficiently low noise characteristics. Kreidberg and Loeb (2016) estimated that the spectral feature of $O_3$ at 9.6 μm of Proxima Centauri b, if present, could be detected in the combined star + planet spectrum after months of observation assuming photon-noise limited precision with a *JWST*-like telescope. Snellen et al. (2017) suggested that the medium-resolution spectrograph (MRS) mode of the Mid-Infrared Instrument (MIRI) onboard *JWST* has a potential to detect 15 μm $CO_2$ feature of Proxima Centauri b in 5 days, using their high-frequency characteristics; such observations will require enabling the time-series observations of MIRI/MRS. Clearly, these observations are contingent on a small noise floor.

## 5. Contextual Information

In this section, we discuss how the contextual information other than the planetary properties may be used to improve the characterization of potentially habitable planets in question. First, we mention the efforts to characterize the host stars (Section 5.1), as one of the most essential ingredients in assessing the planetary environment. We also consider the information of the architecture of the target planetary system (Section 5.2), and the chemical characterization of the gaseous planet(s) in the same system (Section 5.3), both of which will be more easily available than the properties of terrestrial planets.

### 5.1. Properties of the Host Star

#### 5.1.1 Mass, Radius, SED in the Visible/Infrared Range

Precise measurement of planetary mass and radius depends on accurately knowing the host star, and the planetary climate **is** largely affected by the spectral energy distribution (SED) in the visible to near-infrared range. The basic properties of the host stars such as radius, mass, age, and effective temperature may be estimated based on the observed spectra and the distance (estimated from parallax), or near-infrared spectroscopy armed with stellar evolutionary models. Asteroseismology provides additional information to characterize the stellar properties (e.g., Huber et al. 2013), and more asteroseismic data will become available along with searches for transiting planets. In many cases, these basic properties are cataloged with varying accuracy. Depending on the reliability of the model used to derive these values, additional observations of individual host stars may be needed to obtain more precise values for these parameters.

*5.1.2 Activity (SED in UV, X-ray, superflares)*

Photochemical reactions in the planetary atmosphere depend on the SED in the UV range. Photochemistry affects the atmospheric profiles of composition and temperature, influencing the detectability and reliability of some of the biosignatures. For example, model studies (e.g. Segura et al. 2005; Grenfell et al. 2012; Tabataba-Vakili et al. 2016) show that the abundances of $O_3$ and $CH_4$ are sensitive to the UV flux. The UV output may also abiotically produce potential biosignature gases such as $O_2$ (e.g. Hu et al. 2012b; Tian et al. 2014; Domagal-Goldman et al. 2014). In addition, powerful coronal mass ejections, or superflares (Maehara et al. 2012, 2015), would interact with the planetary magnetosphere, and cause energetic particles to flood into the atmosphere and induce in-situ chemical reactions (Airapetian et al. 2016). In particular, Earth-like ($N_2$-$O_2$) atmospheres would form $N_2O$, which could then be a false positive for a biosignature. High-energy radiation towards the X-ray range, called XUV (roughly 1–1000 Å), drives atmospheric loss through thermal mechanisms (Jeans escape and hydrodynamic escape) and non-thermal mechanisms (e.g., through the charge separation driven by XUV ionization; Airapetian et al., 2017). Thus, these high energy fluxes over the history of the star critically impact atmospheric evolution.

The strength of stellar high-energy radiation is related to the magnetic activity (stellar dynamo) (e.g., Noyes et al. 1984), and they are negatively correlated with the age and the spin rotation period (e.g., Wilson 1966, Kraft 1967; Pallavicini et al. 1981; Wright et al. 2011; Astudillo-Defru et al. 2017a). The profile of Ca II H and K lines is used as an observational proxy of the activity (e.g., Wilson 1966, Kraft 1967; Saar, Fischer et al. 2000; Queloz et al. 2001; Wright et al. 2004), while Hα line is also becoming more widely used as an activity tracer for late-type stars (e.g., West et al. 2008; Gomes da Silva et al. 2011; Astudillo-Defru et al. 2017b). With the increasing awareness of its importance for exoplanet study, characterization of high-energy radiation of a wide spectral range of is being advanced using data from the *HST* (UV), *ROSAT*, *XMM-Newton* and *GALEX* (X-ray) and through the development of models to reconstruct the spectra in the wavelength range that is difficult to observe (e.g., Engle and Guinan 2011; Stelzer et al. 2013; France et al. 2013; France et al. 2016; Youngblood et al. 2016; Loyd et al. 2016).

## 5.2. Orbital Architecture of the Planetary System

While we will discover tens to about one hundred Earth-sized planets around HZs in the coming decade, a substantially larger number of giant planets and/or planets in a variety of orbits that are easier to detect will also be discovered. For example, *TESS* will discover >1000 planets with radii larger than $2R_\oplus$ (Sullivan et al. 2015). New ground-based transit surveys, including the Next Generation Transit Survey (Wheatley et al. 2013), will also contribute to unveiling the population of transiting planets. Continuous efforts in radial velocity monitoring will also uncover more planets. In addition, using the astrometry method, the *Gaia* mission (Casertano et al. 2008) is estimated to discover ~21,000 large and distant planets during its nominal 5-year mission (Perryman et al. 2014). Large and distant planets will also continue to be targeted by ground-based direct-imaging observations, as well as by the future coronagraphic instruments on *JWST*, *GMT*, *TMT*, *ELT*, and *WFIRST*. These observations will add a significant number of samples to our catalog of planetary systems whose major architectures (orbits, masses and/or radii of the planets) are known---How is such information related to the future biosignature search?

In general, such information potentially has indirect implications for planets of astrobiological interest through planet formation processes. For example, Earth-sized planets in the HZ of systems with hot Jupiter(s) may be volatile-rich in comparison to Earth, as a consequence of the migration process; specifically, hot Jupiters are believed to form further out in the disk and then subsequently reach short orbital periods through viscous migration, during which process material from beyond the snow line is dragged inward, resulting in small planets with high volatile inventories (Raymond et al. 2006). Likewise, planets in systems whose architectures suggest inward migration (e.g., orbits in resonance) may have also formed beyond the snow line and be volatile-rich (Izidoro et al. 2014). However, planet formation and evolution processes include many uncertainties, and at this point we cannot make definitive predictions for the properties of individual Earth-sized planets, given the variety of compositions that can be produced under similar conditions (Carter-Bond et al. 2012). Conversely, once we have obtained the spectral information of the Earth-sized planets, this will provide, amongst other things, insights into the history of the system.

Another implication from orbital architectures is the effect of other planets in the same system on the long-term climate of the terrestrial planets of interest. Companion planets, particularly giants, will cause a terrestrial planet's orbit and obliquity to evolve (Berger 1978; Laskar & Robutel 1993), especially in the absence of strong tidal forces. Such long-term variations have been shown to be a powerful influence on climate, potentially inducing dramatic changes in global surface temperature, ice/snow cover, and possibly carbon cycling (Hays et al. 1976; Spiegel et al. 2010; Armstrong et al. 2014). Hence, climate modeling efforts which seek to place a potential biosignature into context should not neglect the orbital forcing of additional planets.

## 5.3. Characterization of Larger Planets in the System

While characterization of terrestrial planets will be exceedingly difficult, that of larger, gaseous planets will be more feasible during the coming era, providing us with a rich sample of characterized Jupiter-size to Neptune-size planets. For example, *TESS* and *CHEOPS* transit observations combined with RV measurement will give the accurate radius-mass relationship over a wide range of planetary sizes which will enable the determination of their gas fractions and infer possible formation scenarios. Atmospheric characterization of transiting gaseous planets will be performed with *JWST*, *GMT, ELT*, and the *TMT*. Proposed missions, in particular *FINESSE* and *ARIEL*, will carry out chemical surveys of 500 to 1000 transiting planets, preferentially targeting larger, warmer planets. Spectroscopic observations of phase variations could even provide 3-dimensional atmospheric mapping of gaseous planets, as demonstrated in Stevenson et al. (2014). Complementary to the transit observations, which are biased toward close-in planets, Jupiter-sized planets at distant orbits can be observed by direct imaging. Atmospheric characterization of young distant Jupiter-sized planets through direct-imaging observations have been successfully performed with existing 10-meter class telescopes, and *JWST*, *GMT*, *ELT, TMT*, and *WFIRST* will also be able to perform high-contrast imaging of Jupiter-sized distant planets both in thermal light and in scattered light---If gas giants are present in the same planetary system as the terrestrial planets of astrobiological interest, what can characterization of these larger planets tell us about the habitability of terrestrial planetary companions?

The properties of gas giant atmospheres will provide additional insights into planet formation. For example, the core mass, which may be constrained from the mass-radius relationship, atmospheric properties, and/or the Love number (the value that describes the sensitivity of deformation of a body in response to a tidal force), could indicate the properties of planet-forming region (e.g., Batygin et al. 2009; Nettelmann et al. 2010; Nettelmann et al. 2011); If the atmospheric composition indicates a C/O ratio significantly different from the host star's, it may have formed beyond the snow line (Öberg et al. 2011); Abundance ratios of certain elements may also reflect the composition of accreted planetesimals (Pinhas et al. 2016).

Due to the inherently complex nature of planet formation, however, at this point it is difficult to infer detailed characteristics of the disk from giant envelopes alone. Any further connection to the habitability of terrestrial planets will depend heavily on the robustness of formation and geophysical models (see, e.g., Lenardic & Crowley 2012; Mordasini et al. 2012; Stamenković & Seager 2016; Leconte et al. 2015). However, future development in planet formation and evolution theories may find more direct connections with habitability. Conversely, future characterization of terrestrial planets could provide insights into the formation and evolution pathways of the individual systems, giving useful constraints for the formation models.

## 6. Prospects Beyond 2030

In this section, we explore the possibilities beyond 2030 to further advance our investigations of potentially habitable planets. Section 6.1 is devoted to the introduction of the mission concepts currently being studied at NASA, and Section 6.2 include other ideas that the community has been discussing for far-future projects.

## 6.1. Mission Concepts Currently being Studied in the U.S.

Given the ongoing progress of exoplanet science and the further momentum for the field the near-future missions will provide, there is reason for optimism that an exoplanet mission capable of biosignature detection will be selected for development in the 2020s. In anticipation of the 2020 Decadal Survey, NASA is supporting two community-led mission studies that explore a range of science capabilities, costs, and mission architecture for direct imaging of habitable planets: the *Habitable Exoplanet Imaging Mission* (*HabEx*) and *Large UltraViolet Optical and InfraRed surveyor* (*LUVOIR*). *HabEx* would provide some general astrophysics capabilities, while *LUVOIR* mission would give equal weight to exoplanet imaging and general astrophysics in its design. A third mission study, the *Origins Space Telescope* (*OST*), is considering mid-infrared detection of biosignatures in transit spectra post-*JWST*. All three studies are in the early stages of their mission concept definition with many aspects still under discussion and subject to engineering trade studies. Final architecture reports will be delivered in early 2019.

### 6.1.1 Habitable Exoplanet Imaging Mission (*HabEx*)

The *HabEx* study aims to take the first steps in the search for habitability and biosignatures. It is considering "smaller" telescopes with diameters in the 4-6.5 m size range, unobscured telescope apertures, and some combination of coronagraphs and starshade(s). *HabEx* would be able to search the habitable zones of up to 40 nearby solar-type stars. For a conservative $\eta_\oplus$ of 10%, *HabEx* would find a handful of terrestrial planets for spectroscopic follow-up at wavelengths mostly below 1 μm. Especially at the 4 m aperture size, spectroscopy of these at $\mathbb{R}\sim140$ would require weeks of integration time per target, and rotational brightness modulation would be difficult to detect with fidelity. *HabEx* could provide a substantial science return for comparative planetology, with sub-Neptune-size planets accessible around roughly a hundred to several hundred nearby stars and Jovian planets around roughly a thousand. *HabEx* would carry one general astrophysics instrument, provisionally planned as a UV spectrograph. The *HabEx* engineering design work is led by NASA's Jet Propulsion Laboratory.

### 6.1.2 Large UltraViolet Optical and InfraRed surveyor (*LUVOIR*)

The *LUVOIR* study has the greater ambition to survey a large sample of nearby star HZs in order to constrain the frequency of habitability and biosignatures. *LUVOIR* is baselining larger telescopes with diameters in the 9-15 m size range, providing access to 200-500 HZs, and a "conservative" yield of 20-50 terrestrial planets. Such large target samples are difficult to survey with starshades, thus *LUVOIR* has baselined coronagraphs as its prime starlight suppression architecture. The constraints of existing launch vehicles require that the mission's large telescope aperture be realized with a segmented primary mirror, for which coronagraphy is more technically challenging. *LUVOIR*'s greater collecting area would allow spectra of HZ exoplanets to be made in roughly a day of integration time; the simulated observed data are shown by green points in Figure 6 (Robinson et al. 2011); a subsample of at least a dozen targets could have spectra measured into the near-infrared and have their rotational brightness modulation detected; the comparative planetology would be even richer than in the case of *HabEx*. Terrestrial planets at varying ages in varying orbits will provide insights into how planets have formed and how they evolve over geological time, including the "magma ocean" scenario (Hamano et al. 2015) and the co-evolution of life and the planetary environment (e.g., Kaltenegger et al. 2008).

In addition to its coronagraph, *LUVOIR* would carry four other instruments for general astrophysics. *LUVOIR* is studying whether the general astrophysics camera might be calibrated well enough to provide sub-microarcsecond astrometry, which would enable detection of the stellar reflex motion of HZ terrestrial planets and thus the determination of their masses. The *LUVOIR* engineering design work is led by NASA's Goddard Space Flight Center.

*6.1.3 Origins Space Telescope (OST)*

The *OST* study (Meixner et al. 2016) will characterize the atmospheres of nearby, transiting terrestrial exoplanets using the transmission and emission spectroscopy techniques. One of the primary goals of the *OST* mission is to search for and detect atmospheric biosignatures in multiple systems and assign probabilities to their origins. Like *LUVOIR*, *OST* would be a general astrophysics observatory, but would operate at mid- and far-infrared wavelengths (6 - 600 μm). It is currently baselined to have a segmented primary mirror that is 9 m in diameter and utilizes an off-axis design. It would carry up to five instruments, one of which is being designed specifically to detect biosignatures in exoplanet atmospheres (Matsuo et al. 2016).

In the mid-IR, the main observable is a planet's dayside emission spectrum, as measured using the secondary eclipse technique (Section 3.3). Between 8 μm and 30 μm where the SNR is favorable, there are prominent absorption features due to $CH_4$, $CO_2$, $O_3$, $NH_3$, $N_2O$, and $SO_2$, as well as the $H_2O$ vapor continuum. These features can readily distinguish a wet Earth-like planet from a dry, Venus-like planet with a dense $CO_2$ atmosphere and a Mars-like planet with a thin $CO_2$ atmosphere (Figure 10). The strong $O_3$ band at 9.7 μm allows for the inference of $O_2$, which is a powerful biosignature when combined with other out-of-equilibrium molecular species (such as $CH_4$ at 7.7 μm). Additionally, emission spectroscopy uniquely probes a planet's thermal structure, which is critical towards assessing its habitability.

In addition to measuring planetary emission, mid-infrared observations can take full advantage of a planet's transmission spectrum as measured during primary transit (Section 3.2). Transmission observations place additional atmospheric constraints on the above-mentioned molecules at the planet's terminator. Furthermore, mid-infrared transmission spectra are less affected by the high-altitude aerosols that tend to obscure spectral features at shorter wavelengths (e.g., Hu et al. 2013; Arney et al. 2017).

## 6.2 Ideas for the Far Future

Here we mention some of the more visionary ideas for the far-future missions found in the literature that could further advance our investigations of Earth-sized exoplanets in the search for life. The concepts introduced here are not around the corner in terms of the technological development and funding, and these challenges will not be discussed in detail in this paper.

### 6.2.1 Direct Imaging in the Mid-Infrared

The idea of building a space-based interferometric direct-imaging observatory in the mid-infrared for Earth-sized exoplanets was studied in the proposed mission concepts such as the ESA-led *Darwin* (Léger et al. 1996; Fridlund 2000) and the NASA-led *TPF-I* (Beichman et al. 1999; Lawson et al. 2007), but currently they are not actively studied. While a less challenging contrast of $10^{-7}$ is needed to study habitable exoplanets around solar-type stars in the mid-infrared, going to wavelengths 10-15 times longer than the visible requires telescopes 10-15 times larger (Eq. [9]). For the time being, the telescope dimensions required to study habitable exoplanets in the mid-infrared (> 30 m) can only be realized as interferometers: separate telescopes and a beam combiner distributed on multiple spacecrafts flying in formation. This complexity, combined with the mid-IR requirement to operate at cryogenic temperatures, led the U.S. community to give first priority to architectures operating at visible wavelengths in 2011. Nevertheless, mid-IR high-contrast observations are desirable for characterizing thermal profiles and searching for certain atmospheric molecules of Earth-like planets around solar-type stars, which will be difficult with the near-future instruments. As discussed, signatures in the mid-IR range include various potential biosignatures (e.g., $O_3$, $CH_4$). Thermal emission spectra may also be used to estimate the planetary radius (Des Marais et al. 2002), which otherwise remains unobservable unless it transits. Furthermore, the time variation of thermal emission is affected by planetary obliquity (Gaidos & Williams 2004; Gómez-Leal et al. 2012; Cowan et al. 2012) as well as thermal inertia (Cowan et al. 2012), and thus may be used to make inferences for these parameters.

### 6.2.2. ExoEarth Mapper

Ultimately, we would want a planet imager that has extremely high angular resolution, high enough to spatially resolve the exoplanetary surface (Enduring Quests, Daring Visions (NASA, December 2013), https://science.nasa.gov/astrophysics/documents/). An interferometer in the visible that directly produces, for example, a 10 x 10 pixel map of the surface of an exoplanet, would provide critical, rich information regarding the planetary surface environment. The surface albedo heterogeneity would be directly observed, disclosing the distribution of patchy cloud cover, oceans, and continents. The spatial pattern of clouds would also reveal the atmospheric circulation and possibly imply the underlying topography (e.g. mountains). On compiling a time series, one could also measure the rotation rate and the obliquity of the planet and the seasonality of surface features. We may even search for the spatial distributions of biological surface signatures such as vegetation's red edge and pigments, and their correlation with the distribution of habitats would enhance our confidence level that the detected signatures are indeed biotic. To spatially resolve the disk of an Earth analog at 10 pc distance, at the visible wavelengths, would require an interferometer with baselines of several hundred kilometers.

*6.2.3. Telescope on the Moon*

Like a free-flying space telescope, a telescope on the Moon (e.g., Burns and Mendell 1988) shares the advantage of being outside the Earth's atmosphere and having longer duration access to its targets. The rigid ground of the lunar surface may make the construction of large telescopes and interferometers easier. However, this potential advantage is offset by the disadvantages of large day-night temperature swings, the ~330 hour lunar night, and contamination of optical elements by lunar dust. The lunar far side remains the best site for a low-frequency radio telescope, as it is isolated from terrestrial radio interference.

*6.2.4. 100-meter Class Ground-based Telescope*

The idea of 100-meter class ground-based telescopes was once discussed at the European Southern Observatory (Dierickx & Gilmozzi 2000). The concept for a 74 meter telescope named the Colossus telescope has been proposed (Kuhn et al. 2014). It would likely be two or three decades before ground telescopes larger than ELTs will be pursued.

# 7. Summary: Ideal Timeline

This paper has explored the prospects of future observations to contribute to the general characterization of terrestrial planets in the HZs and to search for biosignatures. Here we summarize them in a serial timeline, for which different aspects were covered in Tables 1 and 2.

Characterization of HZ terrestrial planets in the coming decade will feature transiting planets around late-type (M-type) stars. *TESS* will soon play the primary role in surveying transit signals of nearby short-orbit planets, including Earth-size planets in HZs of late-type stars. *CHEOPS* will then provide measurements of radii down to sub-Neptune size with its ultra-high photometric precision. These planetary systems around nearby late-type stars will allow for RV mass measurements by ground-based high-resolution spectrographs. The set of well-characterized planets in terms of radius and mass will advance the study of the mass-radius relationship of the close-in small planets. Meanwhile, the host stars and the planetary system architecture will be better characterized.

A small number of the discovered transiting potentially habitable planets around nearby late-type stars may be followed up by observations with *JWST,* if the noise floor is smaller than the expected signals. An intensive use of the telescope for a few golden targets will assess whether such planets have atmospheres at all. If atmospheres and signs of habitability are found, a more thorough search for biosigantures may be conducted through transit spectroscopy with JWST or ground-based telescopes. The largest existing ground-based telescope might perhaps access to the most nearby targets,  upgrading high-resolution spectrographs and high-contrast imaging instruments.

ELTs (*GMT*, *ELT*, *TMT*) will start operation in the 2020s, and all of these telescopes are contemplating the instruments for characterization of template terrestrial planets using high-resolution transmission spectroscopy, high-contrast imaging, and the high-resolution high-contrast method. Once such instruments are installed, they will offer invaluable opportunities to detect atmospheric signatures of HZ planets around late-type stars. This characterization of a handful of golden targets is a tremendous near-term opportunity to not just search for life but also test theories, in particular those about the loss and replenishment of atmospheres around terrestrial planets.

Investigation of potentially habitable planets around solar-type (F-, G-, and K- type) stars will be facilitated from the mid-2020's. Around the mid 2020's, HZ terrestrial transiting planets around these stars will be surveyed by *PLATO*. Together with mass measurements from the ground or the TTV method, the mass-radius relationship for these relatively distant planets will be derived; this will provide critical prior knowledge for future directly imaged. With luck, one or two targets may overlap with the nearby targets of future space-based direct imaging missions.

*WFIRST* offers the first possibility to spectrally characterize these HZ planets around solar-type stars. If *WFIRST* is coupled with an external occulter (starshade) to sufficiently block the stellar light to the contrast level $10^{-10}$, it may be able to directly image Earth-sized planets in HZs of solar-type stars, and take low-resolution spectra for crude characterization of their atmospheres. Without an external occulter, *WFIRST* will work with a coronagraph instrument that is expected to operate at contrast limits of about $2 \times 10^{-9}$ at 130 mas separations, enabling for the detection of larger planets, which potentially include terrestrial ones.

There are three options being considered for advancing our search for biosignatures, beyond the initial search that will be conducted over the next ~10 years. The first option is to investigate the same (and similar) targets as observed by *JWST* and/or ELTs (i.e., primarily those around late-type stars) with the higher sensitivity (compared to *JWST*) and expanded wavelength coverage (compared to ELTs') provided by *OST*. The second option, *HabEx*, will target biosignatures on planets orbiting nearby solar-type stars with scattered light spectroscopy, while also enabling follow-up transit spectroscopy of the *JWST*/ELTs targets in the UV to visible wavelengths. The third option, *LUVOIR*, would conduct a survey for biosignatures on planets around hundreds of stars via scattered light spectroscopy of those around solar-type stars as well as detailed follow-up of *JWST*/ELTs targets with transit spectroscopy in the UV to visible wavelengths. These options will allow our community to be responsive to the scientific and technological developments of the next few years. Which option we pursue will be decided by the next US Astrophysics Decadal Survey.

As argued throughout this paper and other papers in this series, finding inhabited planets will not end with the detection of a single feature of a biosignature candidate(s). False positive scenarios must be examined and ruled out based on the environmental context before claims of extraterrestrial life are made. Additional evidence implying habitable conditions will enhance our confidence level for the biological origin of the biosignature candidate. Ultimately, identifying inhabited planets will be the result of successive efforts that accumulate the evidence of the planetary environment, until one finds a set of signatures that cannot be explained by any known abiotic processes *and* can be reasonably explained by evoking the presence of a biosphere. Such efforts should rely on comprehensive characterization of individual planets and the planetary system properties provided by different observations surveyed in this paper, accompanied with the theoretical models of possible varieties in HZ terrestrial planets, both with life and those *without* life.

In this paper, we also discussed the known difficulties in observationally obtaining some of the key parameters to evaluate habitability. Given the limited quality and quantity of the data available in the future, retrievals of the planetary properties will easily suffer from degeneracies that lead to inconclusive or biased interpretations. Thus, developing data analysis techniques and the framework to properly decode available data are essential. It is also important to further explore the methods of characterization, which would correspond to expanding Table 2 and filling in the blanks. Multiple novel ideas were presented in the past decade as reviewed in this paper, and more ideas may be expected to come.

The detection of life across light year distances will perhaps be one of the most difficult measurements ever made, but powerful instruments and careful inquiry should indeed make it possible within the next few decades. No doubt the future will contain hurdles and discoveries that we cannot predict here. We hope, however, that this work will provide a guiding light to steer the way.

## Acknowledgement

We would like to thank the NASA Astrobiology Program and the Nexus for Exoplanet System Science (NExSS) for their support for the NExSS Exoplanet Biosignatures Workshop. Conversations at this workshop, held in the summer of 2016, formed the basis for the drafting of the five review manuscripts in this issue. We also want to thank Mary Voytek, the senior scientist for Astrobiology, for her leadership of NExSS and her feedback on our organization of the workshop and papers. We also thank Natasha Batalha, Mercedes Lopez-Morales, John Fairweather, Avi M. Mandell, Vladimir Airapetian, Sarah Rugheimer, Antonio García Muñoz, and Simon Eriksson, for the feedback during the community input period. We are also grateful to Motohide Tamura for the careful reading and feedback. We acknowledge Jason Tumlinson and Giada Arney for developing the LUVOIR tools used to create Figure 6. The comments from Ramses Ramirez and the other anonymous reviewer greatly improved the clarity of the manuscript. Y. F. is grateful to David S. Amundsen who gave insightful suggestions on the early draft and to Teruyuki Hirano for helpful discussions. R. D. would like to thank Rory Barnes, Victoria Meadows, Rodrigo Luger, Jake Lustig-Yaeger, and Kimberly Bott for enlightening discussions. We thank Lucy Kwok for English proofreading of the manuscript. Y. F.'s research was supported by an appointment to the NASA Postdoctoral Program at the NASA Goddard Institute for Space Studies, administered by Universities Space Research Association under contract with NASA. D.A. acknowledges the support of the Center for Space and Habitability of the University of Bern and the National Centre for Competence in Research PlanetS supported by the Swiss National Science Foundation. R. D. is supported by the NASA Astrobiology Institute's Virtual Planetary Laboratory under Cooperative Agreement number NNA13AA93A. Part of the research was carried out at the Jet Propulsion Laboratory, California Institute of Technology, under a contract with the NASA.

## Author Disclosure Statement

No competing financial interests exist.

**Table 1**. Planned New Observatories

| | expected start | space/ground | aperture | purpose/usage for potentially habitable planets | instruments | wavelength |
|---|---|---|---|---|---|---|
| **TESS** | 2018 (launched) | space | *1 | Discover transiting planets orbiting bright stars | photometry | 0.6-1.0 μm |
| **CHEOPS** | 2018 | space | 32 cm | Provide precise radii of known exoplanets, find transits of RV planets | photometry | 0.4–1.1 μm |
| **JWST** | 2020 | space | 6.5 m | Transmission/eclipse spectroscopy, Phase curves | spectroscopy (NIRISS, NIRSpec, NIRCam, MIRI) | 0.6-28.5 μm |
| **GMT** | 2023 | ground | 24.5 m | Transmission spectroscopy<br>High-contrast imaging<br>High-contrast imaging with High-resolution spectroscopy | spectrogscopy, coronagraphy | 0.3- μm? |
| **ELT** | 2024 | ground | 39.3 m | | | |
| **TMT** | 2027 | ground | 30 m | | | |
| **PLATO** | 2026 | space | *2 | Discover and characterize transiting planets around bright stars, including planets in HZs of solar-type stars | photometry | 0.5-1.05 μm |
| **WFIRST** | 2025 | space | 2.4 m | High-contrast imaging<br>(+ Discover planets by microlensing) | coronagraphy, low-resolution spectroscopy | 0.6-0.95 μm (CGI) |

*1 4 cameras with 10.5 cm aperture each
*2 26 cameras with 12 cm aperture each

**Table 2**: Expected timeline of characterizing planetary parameters. Each cell (surrounded by thick lines) is divided into 2x2 sub-cells; The upper-left and upper-right sub-cells present the methods that can be applied only to transiting planets around late-type stars and solar-type stars, respectively, while the lower sub-cells show the methods that are not limited to transiting planets. Prospects of *HabEx*, *LUVOIR* and *OST* are grayed because not all of them will be realized.

| property | Examples of possible inferences | | -2020 | | 2020-2030 | | 2030- | | visionary |
|---|---|---|---|---|---|---|---|---|---|
| | | | late-type stars | solar-type stars | late-type stars | solar-type stars | late-type stars | solar-type stars | |
| **mass** | - | transit | TTV? | | TTV? | | | | |
| | | general | RV (ground) | RV? (ground) | | | | astrometry (LUVOIR) | |
| **radius** | - | transit | transit (K2, TESS, CHEOPS, ground) | | transit (PLATO) | | | | |
| | | general | | | | | | | HCI in MIR? |
| **transmission spectra** | atmospheric composition, haze/clouds | transit | | | transmission (JWST) HRS transmission (ELTs) | | transmission (OST/LUVOIR/ HabEx) | | |
| | | general | | | | | | | |
| **Scattered-light spectra** | atmospheric/ surface composition, surface pressure, spin rate, obliquity, clouds | transit | | | | | | | |
| | | general | | | HRS+HCI? (ELTs) HCI? (ELTs) | HCI? (WFIRST with starshade) | HCI (LUVOIR/ HabEx) | | Planet Mapper in VIS/NIR? |
| **thermal emission spectra** | atmospheric composition, thermal structure, clouds | transit | | | secondary eclipse (JWST) | | secondary eclipse (OST) | | |
| | | general | | | phase curves (JWST) | | phase curves (OST) | | HCI in MIR? |

RV: radial velocity, HCI: high-contrast imaging (in the visible/near-infrared unless otherwise noted), HRS: high-resolution spectroscopy, TTV: transit timing variation, VIS: visible, NIR: near-infrared, MIR: mid-infrared

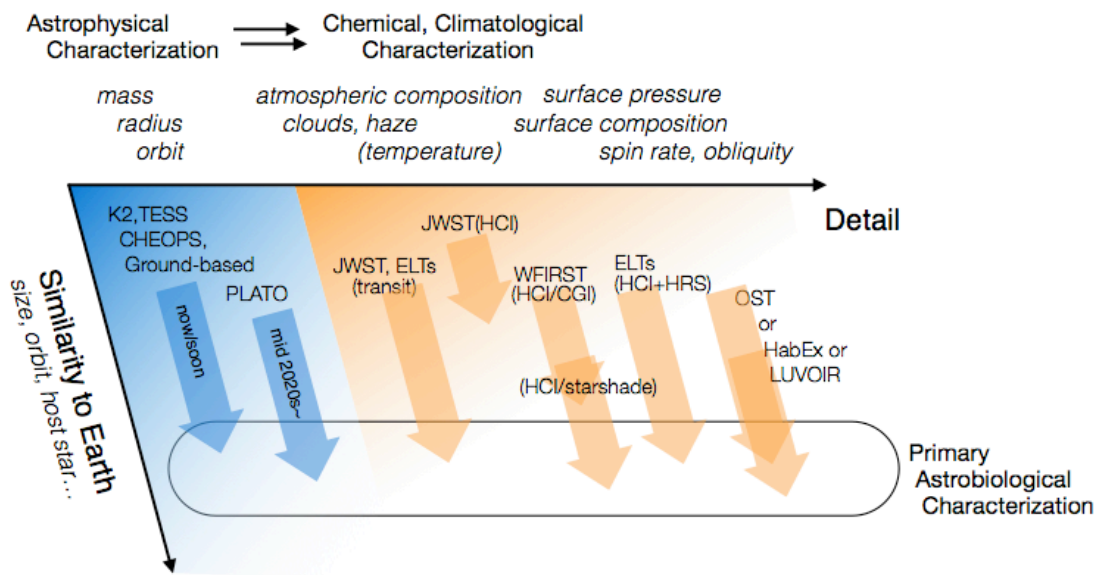

**Figure 1.** Schematic figure showing astrophysical, chemical, climatological, and astrobiological characterizations and the possible contributions from the current and future missions (HCI: high-contrast imaging, HRS: high-resolution spectroscopy).

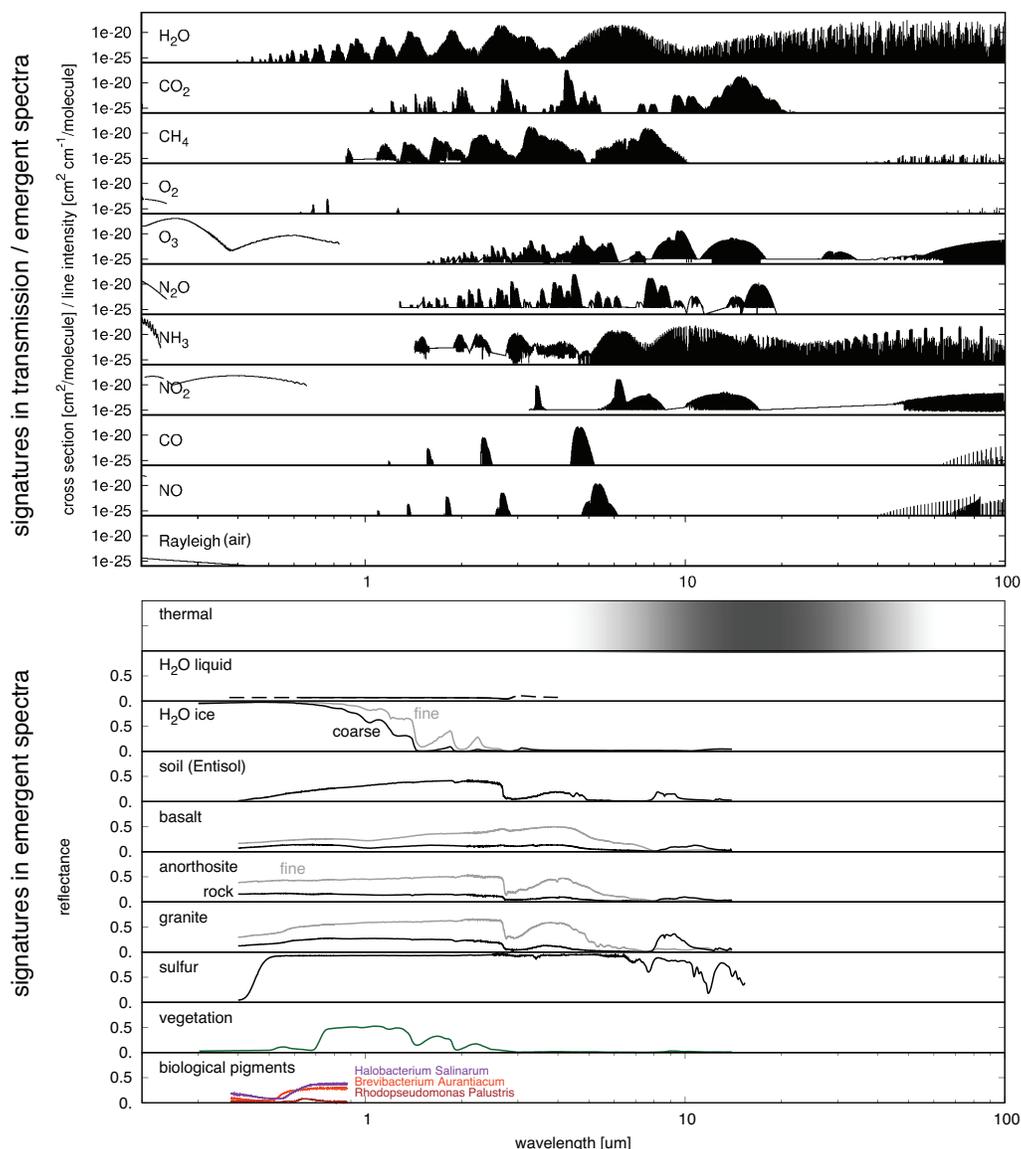

**Figure 2**. Examples of atmospheric and surface spectral features of temperate terrestrial planets. <u>Upper panel</u>: atmospheric signatures, which can in principle be probed through both transmission spectra and emergent spectra. The continuous features of molecules at shorter wavelengths are absorption cross section of molecules at approximately 300K, taken from MPI-Mainz UV/VIS Spectral Atlas of Gaseous Molecules of Atmospheric Interest (Keller-Rudek et al., 2013), shown in log scale from $10^{-26}$ to $10^{-16}$ cm$^2$ / molecule. Original data sources are Yoshino et al., (1988) for $O_2$, Brion et al., (1998) for $O_3$, Selwyn et al., (1977) for $N_2O$, Cheng et al., (2006) for $NH_3$, Merienne et al., (1997), Coquart et al., (1995), and Vandaele et al., (1998) for $NO_2$. The lines at longer wavelengths are line intensity at 296K and 1 atm in log scale from $10^{-26}$ to $10^{-16}$ cm$^2$ cm$^{-1}$ /molecule taken from HITRAN2012 database (Rothman et al., 2013). <u>Lower panel</u>: thermal radiation and reflectance spectra of surface materials, which can be probed in emergent light. The reflectance is shown in linear scale from 0 to 1. All data but biological pigments are taken from ECOSTRESS Spectral Library (Baldridge et al., 2009; Meerdink et al., in prep.). The data of biological pigments are from VPL spectral databases (Schwieterman et al., 2015).

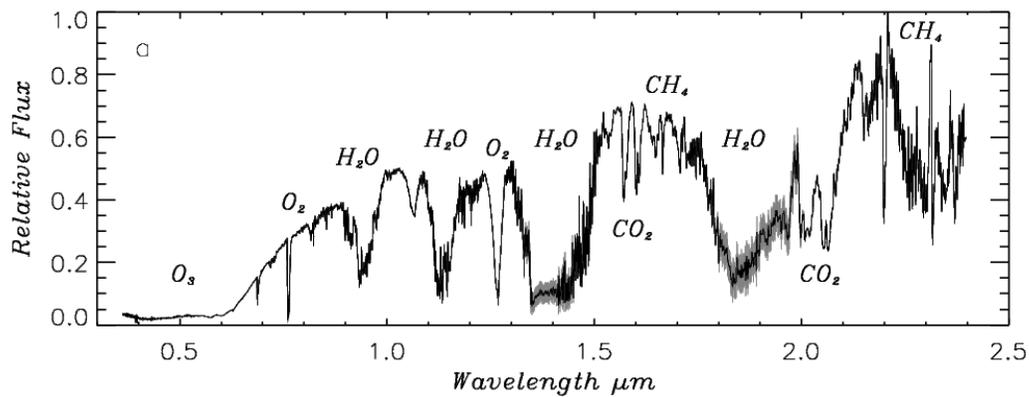

**Figure 3.** Transmission spectra of the Earth observed at the lunar eclipse, taken from Pallé et al. (2009). The spectral resolution is ℝ~960 in the optical and ℝ~920 in the near-infrared.

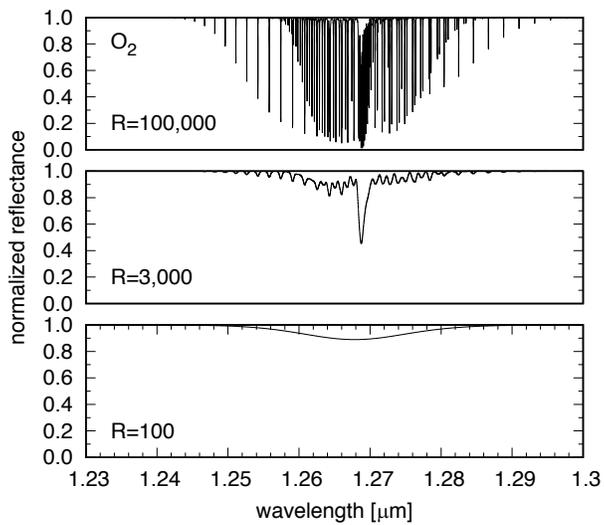

**Figure 4**. O₂ 1.27 μm features in reflectance spectrum of Earth atmosphere with varying spectral resolution.

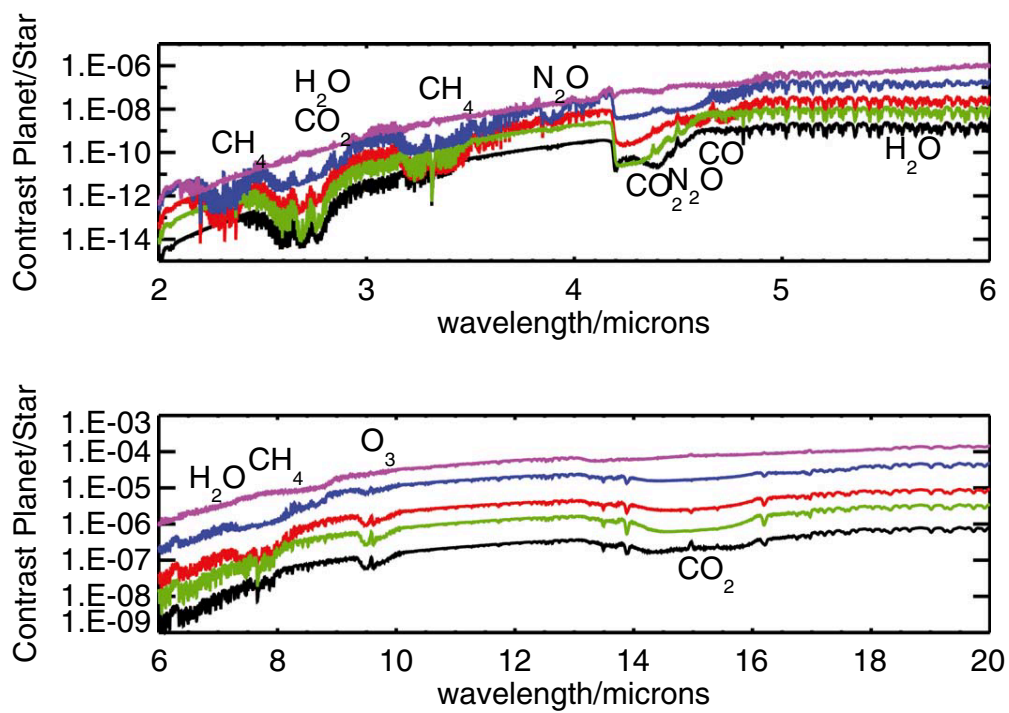

**Figure 5**. Modeled thermal emission spectra of cloud-free Earth-like planets around the Sun (black), AD Leo (red), an M0 star (green), an M5 star (blue), and an M7 star (magenta), taken from Rauer et al. (2011). Reproduced with permission © ESO.

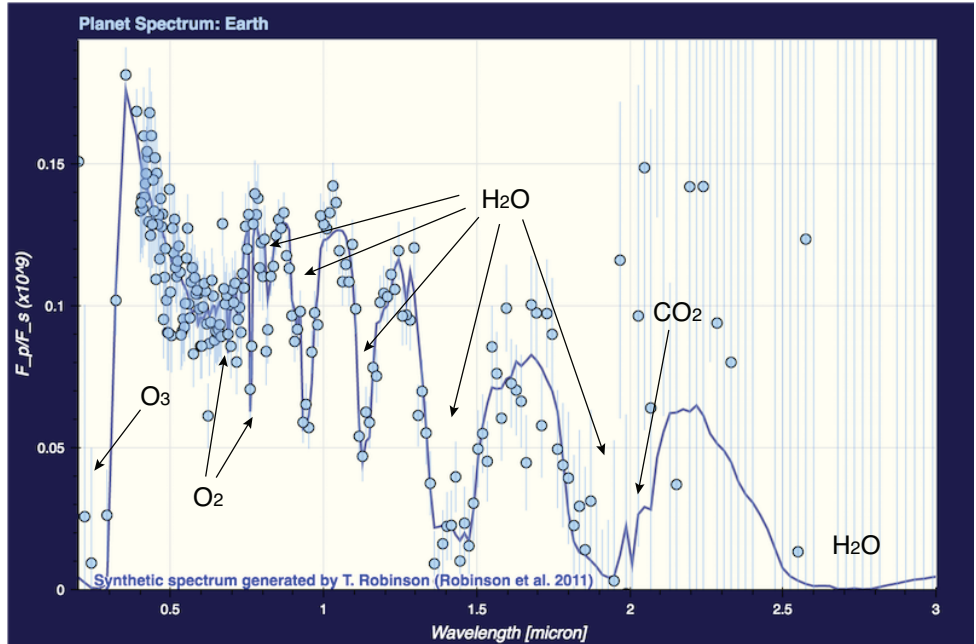

**Figure 6.** A modeled scattered light spectrum of the Earth (blue solid line) and the mock observation of an Earth-twin at 5 pc away assuming a LUVOIR-type telescope with 12 meter diameter and 24 hours of integration time, with resolution $\mathbb{R} = 150$ (blue points with error bars), generated at http://jt-astro.science:5106/coron_model. The theoretical line and the noise model are based on Robinson et al. (2011) and Robinson et al. (2016), respectively.

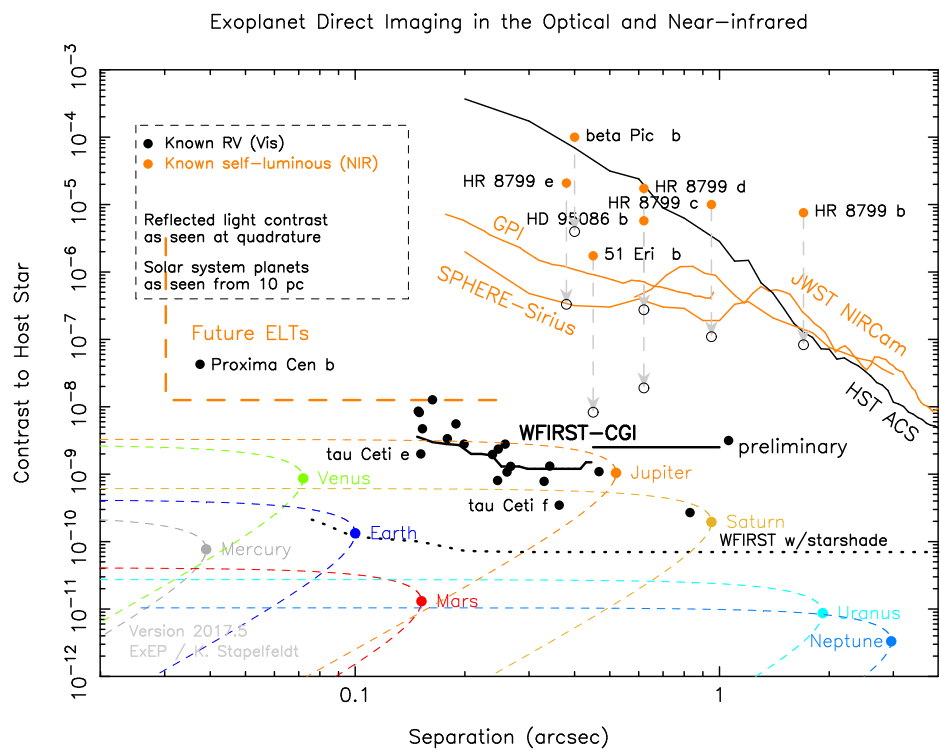

**Figure 7**. The star-planet contrast and the star-planet separation of known planetary systems (points), and the performance of existing and future high-contrast imaging instruments (lines). This is the October 2017 version of a plot maintained by the NASA Exoplanet Exoploration Program Office. The orange points correspond the near-infrared brightness of known self-luminous directly-imaged planets, while the open circles show their theoretical I-band contrast. The black points show the theoretical V-band contrast of planets detected by the RV method. The Solar-system planets at 10 pc at the maximum separation are presented in colored points; the dashed lines from these points indicate their orbital phase variations as seen from a direction inclined 30 degrees from the ecliptic. The self-luminous planets detected to date are at contrasts of $10^{-6}$ and brighter, while $10^{-9}$ contrasts is needed to detected Jupiter in scattered light and $10^{-10}$ detect the Earth as seen from outside. The data sources for the instrumental performance lines are as follows: The JWST NIRCam and HST ACS curves were provided by John Krist for Lawson et al. (2012). The GPI curve is for H band and provided by Bruce Macintosh (personal communication). The SPHERE-Sirius curve is for K band (Vigan et al. 2015, Figure 2). The 2017 WFIRST-CGI curve was provided by deputy instrument scientist Bertrand Mennesson (personal communication). The starshade curve is from Stuart Shaklan (personal communication). The performance curves shown for it are preliminary as of October 2017 and subject to revision.

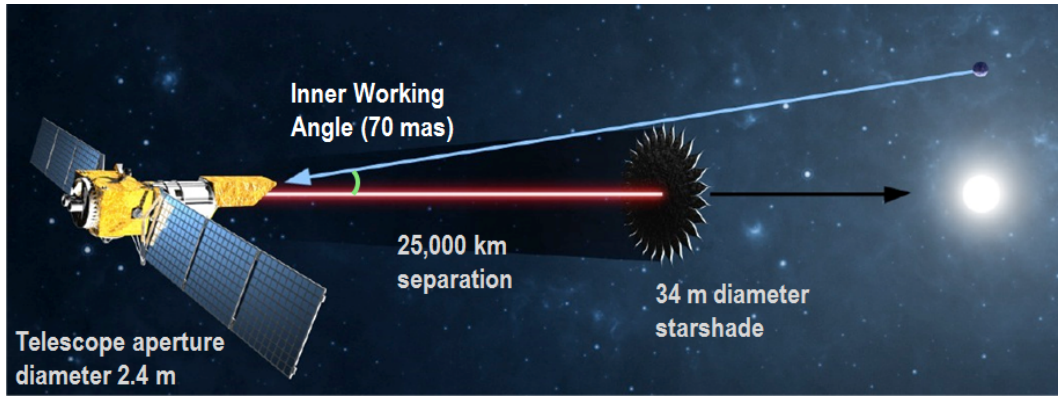

**Figure 8**. Flight configuration of a 34 m starshade in exo-Earth search mode (bandpass of 425-605 nm) was studied as part of the Exo-S Extended Study (https://exoplanets.nasa.gov/internal_resources/225/; Seager et al., 2015). This particular configuration attempts to maximize exo-Earth yield and assumes a 3 year mission; other options considered included both larger and smaller starshades with shorter and longer mission lifetimes.

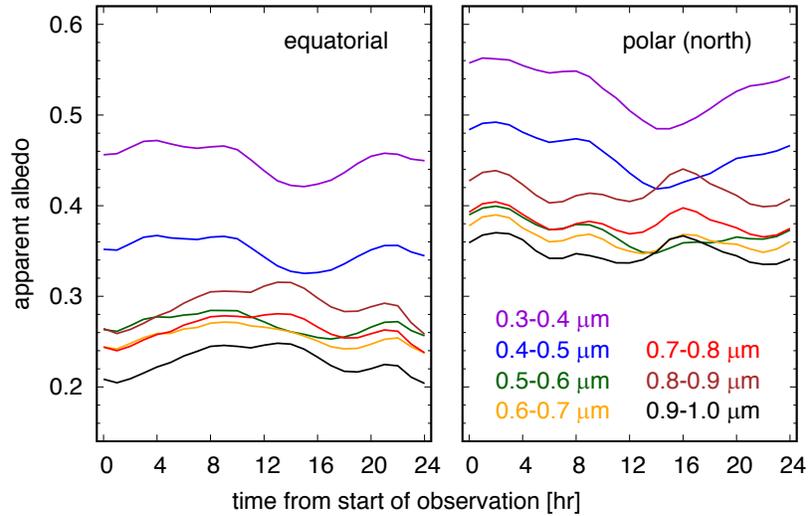

**Figure 9**. Seven-band diurnal light curves of the disk-integrated scattered light of Earth, obtained with the EPOXI mission (Cowan et al., 2011; Livengood et al., 2011). <u>Left panel</u>: The equatorial observation started on March 18, 2008, with phase angle 57 degree. <u>Right panel</u>: The north-polar observation started on March 27, 2009, with phase angle 87 degree.

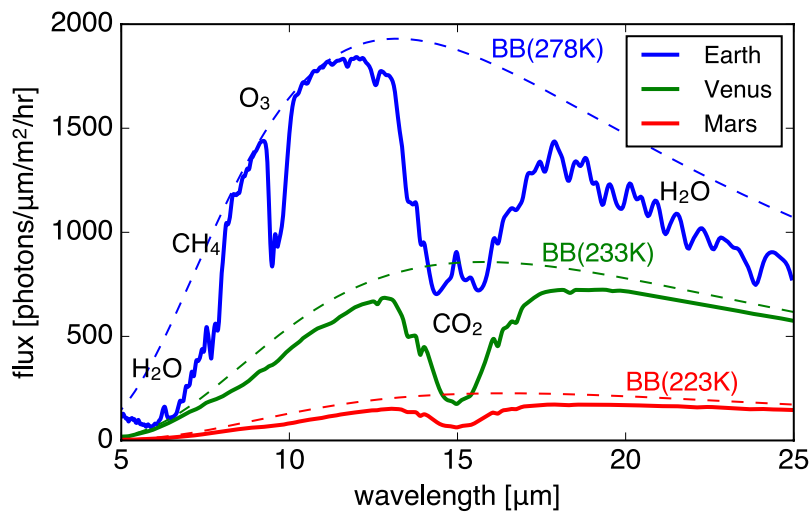

**Figure 10**. Solid lines: Thermal emission spectra of Earth, Venus, and Mars. The Earth spectrum is from Robinson et al., (2011). The spectra of Venus and Mars were modeled using the radiative transfer code SMART, assuming the 1D atmospheric profiles of each planet. Venus data is from Giada Arney, and Mars data is from Robinson & Crisp (2018). Dashed lines: Black body emission from a planet of the same radius with the approximately maximum brightness temperature of each planet in this range. See also Selsis et al. (2008) and Kaltenegger (2017).